\documentclass[12pt]{article}

\usepackage{amsfonts} 
\usepackage{amsmath}
\usepackage{amsthm}
\usepackage[usenames,dvipsnames]{color} 
\usepackage[english]{babel}

\usepackage{amssymb}

\usepackage{bm}
\usepackage[inner=3cm,outer=3cm,bottom=3cm,top=3cm]{geometry}
\usepackage{latexsym,pdfsync,xcolor,graphicx}
\usepackage{array}
\usepackage{subfig}
\usepackage{tikz}
\usepackage{natbib}
\usepackage{appendix}

\usepackage{hyperref}

\usepackage{subfig}
\usepackage{enumerate} 
\usepackage{multirow} 
\usepackage{rotating}
\usepackage{float}
\usepackage{textcomp}
\usepackage{booktabs}
\usepackage{lscape}
\usepackage{titling}

\numberwithin{equation}{section}
\theoremstyle{plain}

\newtheorem{prop}{Proposition}[section]

\begin{document}

\title{\large Nonparametric tests for circular regression} 

\author{\small Mar\'ia Alonso-Pena$^{a}$, Jose Ameijeiras-Alonso$^{b}$ and Rosa M. Crujeiras$^{a}$}
\date{}


\maketitle
\vspace{-1.8cm}
\footnotesize
\begin{center}
$^a$ Department of Statistics, Mathematical Analysis and Optimization, Universidade de \\Santiago de Compostela.\\
$^b$ Department of Mathematics, KU Leuven
\end{center}
\vspace{0.3cm}

\hrule
\vspace{0.2cm}
\normalsize
\noindent
\textbf{Abstract}\\
No matter the nature of the response and/or explanatory variables in a regression model, some basic issues such as the existence of an effect of the predictor on the response, or the assessment of a common shape across groups of observations, must be solved prior to model fitting. This is also the case for regression models involving circular variables (supported on the unit circumference). In that context, using kernel regression methods, this paper provides a flexible alternative for constructing pilot estimators that allow to construct suitable statistics to perform no-effect tests and tests for equality and parallelism of regression curves.  Finite sample performance of the proposed methods is analyzed in a simulation study and illustrated with real data examples.
\vspace{0.1cm}

 
\vspace{0.1cm}

\hrule

\section{Introduction}
\label{sec:intro}

Regression methods provide a classical approach for modeling the dependence relationship between two variables. Many different models have been proposed over the years, considering both parametric and nonparametric approaches as well as including adaptations to more complex settings beyond the euclidean case.  A particular situation where the usual regression models cannot suitably handle is the one where the response and/or the covariate can be expressed as angles on the unit circumference, i.e., \textit{circular data}. For this type of observations, the periodicity and the nature of the support hampers the use of linear statistical methods (i.e. tools designed for real-valued random variables) even for a simple descriptive analysis.

Just to illustrate the regression ideas in a circular context, let us consider two different real data examples. The first dataset is given in \cite{Anderson-Cook1999} and contains mechanical measurements on flywheels. A flywheel is a device designed to regulate an engine's rotation. It is a heavy wheel attached to a rotating shaft and it is used to store rotational energy in an efficient way. Balancing flywheels is crucial in vehicles production in order to ensure that the rotation transmits minimal vibration. When correcting the balance, one obtains an angular component measuring the angle of imbalance and a linear component evaluating the magnitude of the correction required to balance the flywheel. Modeling the relationship between the angle and the magnitude of correction can be helpful for a better understanding of the process, leading to the minimization of the costs by creating more efficient designs. The data given in  \cite{Anderson-Cook1999} contains measurements of the angles of imbalance of 60 flywheels, as well as the measurements of the corrections required (in inch-ounces). A circular representation of the data is given in the left panel of Figure~\ref{fig:data_anderson}.

Our second example was obtained from an experimental study described in \cite{Scapinietal2002} and further analyzed by \cite{Marchetti_Scapini2003}, where the authors investigate the direction of movement of a group of sand hoppers of the species Talitrus saltator under natural conditions. To record the data, two different circular arenas with cross traps placed at the circumference were used. The animals were released in the arenas, and once they made an orientation choice they were caught in one of the traps, which were separated from each other by an angle of 5\textdegree.  In addition, other variables were recorded, such as the temperature (linear) and the sun azimuth (circular). Figure~\ref{fig:data_sandhoppers} shows a representation on the cylinder of the direction of the animals with respect to the temperature (top row), whereas the escape direction with respect to sun azimuth is plotted over a torus (bottom row). 

For modeling both datasets, parametric and nonparametric methods can be considered depending on the desired flexibility of the model. A review on parametric circular regression methods can be found in \cite{Jammalamadaka_SenGupta2001}. In what follows, the main parametric ideas are briefly reviewed. First, the flywheels dataset is an example where a regression model with linear response and circular covariate  (\textit{circular-linear regression}) could be useful, where the regression function is defined in the surface of a cylinder. In that case, in a similar approach to the linear models, the effect of the predictor can be accounted through its sine and cosine components (see, e.g., \cite{Mardia_Sutton1978}).

When measuring the relation of the direction of sand hoppers (circular response) with respect to the temperature (real-valued predictor), the \textit{linear-circular regression} function can also be regarded to lie on the surface of a cylinder. For modeling the dependence between these two variables, it is usually assumed that the responses follow a specific parametric distribution, where the circular mean of the distribution is modeled as a function of the predictor. Specifically, \cite{Fisher_Lee1992} assume that the response variable follows a von Mises distribution (see equation~(\ref{eq:vonMises})) with constant concentration, and the covariate directly affects the location parameter, via a link function. The same authors also consider other models accounting for a possible effect of the covariate over the concentration. On the other hand, \cite{Presnelletal1998} avoid the selection of the link function by considering a projected model from a bivariate normal distribution.

The last scenario, the direction of the sand hoppers depending on the sun azimuth, involves two circular variables (\textit{circular-circular regression}) and can be represented on the surface of a torus. \cite{Jammalamadaka_Sarma1993} introduce polynomial models on sine and cosine components of the response, defined over sine and cosine components of the covariate for this setting.

Despite the feasible direct application of the previous ideas to our datasets, these parametric models might not be flexible enough to capture the features of the regression functions. More flexible approaches are found in nonparametric methods. In the circular-linear context, \cite{DiMarzioetal2009} derived a kernel type estimator for circular predictors by using circular kernel functions. For the cases where the responses are circular, \cite{DiMarzioetal2012} proposed a nonparametric estimator for the regression function. A recent overview of nonparametric directional regression can be found in \cite{Ley_Verdebout2017}.

The pursued data-driven character of kernel methods makes it difficult to ascertain which features of the estimation correspond to the underlying regression function and which ones are just sample noise. Hence, a first question to answer before proceeding with a regression approach is to actually verify if the covariate has a significant effect on the response. With that objective, a \textit{no-effect} test is provided in this paper.

Another interesting problem arises when a discrete variable determining different groups for the observations is considered. In the flywheels example, the metallic molding employed in the production process is made out of four different metals, dividing the observations into four groups. As for the sand hoppers data, one of the arenas used allowed the view of both the sky and the landscape, while in the other, the landscape was screened off, so that only the sky was visible. Therefore, the variable indicating the type of arena determines two different groups of observations. In this context, it is interesting to assess if the curves, for each group, are the same (\textit{equality test}) or if the distance between them is constant (\textit{parallelism test}).

In this manuscript we present new proposals to overcome these problems in the different regression models involving a circular response and/or covariate. Nonparametric no-effect tests were introduced by \cite{Bowman_Azzalini1997} in the linear context, assuming normal and homoscedastic errors and approximating the distribution of the test statistic by a shifted and scaled $\chi^2$ distribution. Analysis of Covariance (ANCOVA) models were introduced also in the linear context by \cite{Young_Bowman1995}, under the same assumptions for the residuals as in the no-effect test. The authors present two different tests to investigate the equality and parallelism of the curves across different groups. The proposals presented in this manuscript extend the no-effect and the ANCOVA tests to the three different contexts of circular regression.

This paper is organized as follows. Section~\ref{sec:regression} provides some background on nonparametric regression models involving circular variables (as covariates and/or responses), introducing a no-effect test. In Section~\ref{sec:ancova}, the ANCOVA regression models involving a circular response and/or covariate are presented, jointly with the testing proposals for assessing equality and parallelism of the regression curves. The finite sample performance of the tests is analyzed in Section~\ref{sec:simulation}. In Section~\ref{sec:data}, the practical use of the proposals is illustrated with the flywheels and the sand hoppers examples. Final conclusions are reported in Section~\ref{sec:conclusions}.

\section{Some background on regression models with circular variables}
\label{sec:regression}

In the first place, this section is aimed to review different proposed nonparametric regression models. Through this section we will denote real-valued responses by $Y$, circular responses by $\Phi$, circular predictors by $\Theta$ and $\Delta$ will denote a general covariate, which may be real-valued or circular. Sample individuals identified by $j=1,\ldots,n$, where $n$ is the total sample size. 

New proposals for no-effect tests are given in the second part of the section. In particular, we adapted the ideas of \cite{Bowman_Azzalini1997} to the case having a circular predictor and real-valued response. For the case circular response scenario, an adaptation of the proposal for euclidean variables is not longer valid and, thus, we propose a new bootstrap procedure for testing the hypothesis of no-effect.

\subsection{Nonparametric regression}                  
\label{subsec:nonparametric}

\subsubsection{Circular covariate and real-valued response}

The relationship between a circular predictor variable and a real-valued response variable, given a bivariate sample of both variables, may be described as 
\begin{equation}
Y_j=m(\Theta_j)+\varepsilon_j,
\label{eq:nonpar_circular_linear}
\end{equation}
where $\varepsilon_j$ are iid errors with zero mean and finite standard deviation $\sigma$. Regarding the estimation of the regression function, \cite{DiMarzioetal2009} consider a local trigonometric polynomial fit $\beta_0+\beta_1 \sin(\Theta_j - \theta)$, where $\beta_0\equiv \beta_0(\theta)$ is the regression function and $\beta_1\equiv \beta_1(\theta)$ corresponds to the first derivative of $m$. The local parameters $\beta_0$ and $\beta_1$ are estimated via weighted local least squares, where the weights are given by a circular kernel $K_\kappa$. Through this paper, this kernel is taken as a von Mises density, with zero mean direction and concentration parameter $\kappa$,
\begin{equation}
K_\kappa(\theta) = \frac{e^{\kappa\cos(\theta)}}{2\pi I_0(\kappa)}, \quad \mbox{ where } \theta\in [0,2\pi)
\label{eq:vonMises}
\end{equation} 
and $I_0(\kappa)$ is the modified Bessel function of the first kind and order 0. The (circular) concentration $\kappa$ plays the opposite role of the (linear) bandwidth $h$, in the sense that large value of $\kappa$ normally leads to undersmoothed estimations. For each $\theta \in [0,2\pi)$, the weights given to each observation $\Theta_j, \ j=1,...,n,$ will depend on the distance to the fixed point $\theta$. Thus, the estimated curve at a fixed $\theta$ will be $\hat{m}(\theta)=\hat{\beta}_0$, where 

\begin{equation}
(\hat{\beta}_0,\hat{\beta}_1)=\arg \min_{(a,b)} \sum_{j=1}^{n}K_\kappa(\theta-\Theta_j)[Y_j-(a+b\sin(\theta-\Theta_j))]^2. 
\label{eq:nonpar_fit}
\end{equation}

\subsubsection{Circular response}
Given an angular response $\Phi$ and a predictor $\Delta$, either real-valued or circular, the regression model is given by
\begin{equation}
\Phi_j=[m(\Delta_j)+\varepsilon_j](\mbox{mod }2\pi),
\label{eq:nonpar_linear_circular}
\end{equation}
where $\varepsilon_j$ are iid random angles with zero mean direction and finite concentration. Consider the following circular distance between two angles $\Theta$ and $\Psi$, defined as
\begin{equation}
d(\Theta,\Psi)=1-\cos(\Theta-\Psi).
\label{eq:circ_distance}
\end{equation}
In order to minimize the risk associated to $d(\Phi,m(\Delta))$, \cite{DiMarzioetal2012} propose estimating $m$ as
\begin{equation}
\hat{m}(\delta)=\mbox{atan2}(\hat{g}_1(\delta), \hat{g}_2(\delta)),
\label{eq:circ_mestimator}
\end{equation} 
with $\mbox{atan2}$ returning the angle between the $x$-axis and the vector from the origin to $(\hat{g}_1(\delta),\hat{g}_2(\delta))$, where			
\[\hat{g}_1(\delta)=\frac{1}{n}\sum \sin(\Phi_j)W(\Delta_j-\delta), \quad \hat{g}_2(\delta)=\frac{1}{n}\sum \cos(\Phi_j)W(\Delta_j-\delta).\]
Different ways of estimating the linear or circular weights $W(\cdot)$ can be chosen. In this paper, the circular analogue of the local linear weights is considered. Thus, the weights $W(\Delta_j-\delta)$ are equal to one of the following quantities depending on the nature of the predictor,

{\footnotesize
	\begin{eqnarray}
	&\frac{1}{n}L_h (X_j-x)\left[ \sum_{k=1}^{n}L_h (X_k-x)(X_k-x) - (X_j-x)\sum_{k=1}^{n}L_h (X_k-x)(X_k-x) \right], \label{lllinear}\\
	&\frac{1}{n}K_\kappa (\Theta_j-\theta)\left[ \sum_{k=1}^{n}K_\kappa (\Theta_k-\theta)\sin^2(\Theta_k-\theta) - \sin(\Theta_j-\theta)\sum_{k=1}^{n}K_\kappa (\Theta_k-\theta)\sin(\Theta_k-\theta) \right]. \label{llcircular}
	\end{eqnarray}}
In equation~(\ref{lllinear}), the predictor $\Delta$ is linear, thus $\Delta$ is replaced by $X$ (real-valued variable), $\delta$ by $x\in\mathbb{R}$ and $L_h$ is a linear kernel. In particular, the Gaussian kernel is considered in this paper, i.e., $L_h$ is the Gaussian density, with zero mean and standard deviation $h$. While in equation~(\ref{llcircular}), a circular predictor $\Delta$ is considered, thus $\Delta$ is replaced by $\Theta$ (circular variable), $\delta$ by $\theta\in[0,2\pi)$, and $K_\kappa$ is a circular kernel with zero mean and concentration $\kappa$. In this paper, the von Mises kernel is employed. 

\subsection{A no-effect test}
\label{subsec:noeffect}
As mentioned in the Introduction, a first question to analyze when trying to fit a regression model is to assess if there is a significant effect of the covariate over the response. For that purpose, nonparametric no-effect tests will be proposed for the different regression scenarios involving a circular response and/or covariate.

\subsubsection{Real-valued response and circular covariate}
Consider the regression model in (\ref{eq:nonpar_circular_linear}). A test to ascertain the effect of the covariate is constructed with the following hypotheses:
\begin{eqnarray*}
	H_0&:& Y_j=\gamma + \varepsilon_j, \quad \gamma \in \mathbb{R},\\
	H_1&:& Y_j=m(\Theta_j)+ \varepsilon_j,  \quad m(\Theta_j)\neq\gamma \ \mbox{for some} \ j\in \{1,...,n\}.
\end{eqnarray*}
First, we will assume that the errors follow a normal distribution with mean zero and variance $\sigma^2$, although this condition will be relaxed later.  A test statistic can be constructed by adapting the ideas by \cite{Bowman_Azzalini1997} to the circular context, using the nonparametric estimator derived from (\ref{eq:nonpar_fit}). Therefore, the residual sums of squares are used to quantify how much the models explain the data under each of the two hypotheses. Then, the test statistic takes the form
\[C_1=\frac{RSS_0-RSS}{RSS},\]
where the residual sums of squares under the null and the alternative are given by
\[ RSS_0=\sum_{j=1}^{n}(Y_j-\hat{\gamma})^2, \quad \mbox{and} \quad RSS=\sum_{j=1}^{n}(Y_j-\hat{m}(\Theta_j))^2.\]
The constant parameter $\gamma$ is estimated with the sample mean of the responses, while the regression curve under $H_1$ is estimated with the nonparametric estimator for circular predictors and real-valued responses ($\hat{\beta}_0$ in (\ref{eq:nonpar_fit})). The nonparametric estimator is a linear form in the data, i.e.,
$ \bm{\hat{m}}=\bm{SY}, $
where $\bm{\hat{m}}$ is the vector with the fitted values, $\bm{S}$ is the smoothing matrix and $\bm{Y}$ is the vector containing the responses. Consequently, the residual sums of squares can be expressed in vector-matrix notation
$$ RSS_0=\bm{Y}'(\bm{I}_n-\bm{L})'(\bm{I}_n-\bm{L})\bm{Y} \quad \mbox{and} \quad   RSS=\bm{Y}'(\bm{I}_n-\bm{S})'(\bm{I}_n-\bm{S})\bm{Y}, $$
where $\bm{L}$ is a $n\times n$ matrix with $n^{-1}$ in all its components and $\bm{I}_n$ is the identity matrix of order $n$. Thus, the test statistic can be rewritten as
$$ C_1=\frac{\bm{Y}'\bm{BY}}{\bm{Y}'\bm{AY}}, $$
with $\bm{A}=(\bm{I}_n-\bm{S})'(\bm{I}_n-\bm{S})$ and $\bm{B}=\bm{I}_n-\bm{L}-\bm{A}$. Now, a $p$-value for the test is obtained as
$$ p=\mathbb{P}\left(\frac{\bm{Y}'\bm{BY}}{\bm{Y}'\bm{AY}}>Obs\right)=\mathbb{P}(\bm{Y}'(\bm{B}-\bm{A}\cdot Obs)\bm{Y}>0),$$
with $Obs$ being the observed value of the statistic. Although under the null hypothesis $\mathbb{E}(Y_j)=\gamma$, it is easy to see that $\gamma$ disappears in the expression of $C_1$ due to the differences involved. Then, the $p$-value calculation is equivalent to
$$ p=\mathbb{P}(\bm{\varepsilon}'(\bm{B}-\bm{A}\cdot Obs)\bm{\varepsilon}>0). $$
Now, given that the  matrix $\bm{B}-Obs\cdot\bm{A}$ is  symmetric, we have that $\bm{\varepsilon}'(\bm{B}-\bm{A}\cdot Obs)\bm{\varepsilon}$ is a quadratic form in normal variables of the type $\bm{z'Cz}$ where $\mathbb{E}(\bm{z})=\bm{0}$ and $\bm{C}$ is symmetric. Then, the results in \cite{Bowman_Azzalini1997} can be applied, approximating the distribution of $C_1$ by a more convenient one. With that objective, note that the first three cumulants of $\bm{\varepsilon}'(\bm{B}-\bm{A}\cdot Obs)\bm{\varepsilon}$ can be obtained as
$$ \nu_s=2^{s-1}(s-1)!\mbox{tr}{(\bm{VC})^s},\quad s=1,2,3, $$
where $\mbox{tr}$ denotes the trace operator and $\bm{V}=\bm{Cov}(\bm{z},\bm{z})$. Then, the distribution of $\bm{\varepsilon}'(\bm{B}-\bm{A}\cdot Obs)\bm{\varepsilon}$ is approximated by a shifted and scaled $\chi^2$, with parameters calculated as 
\[a=|\nu_3|/(4\nu_2), \quad b=(8\nu_2^3)/\nu_3^2, \quad c=\nu_1-ab,\]
with $a$ being the scale parameter, $c$ being the location parameter and $b$ the number of degrees of freedom. Thus, the $p$--value can be computed as $\mathbb{P}\left[\chi_{b}^{2} > -c / a\right]$.\\

Note that the Gaussian condition about the errors can be relaxed by assuming that the errors have zero mean and constant variance. Then, the calibration of the test can be done by bootstrap, similarly to the other scenarios to be presented next.

\subsubsection{Circular response}
Consider now the regression model in (\ref{eq:nonpar_linear_circular}). The following hypotheses are used to determine the significance of the predictor variable:
\begin{eqnarray*}
	H_0&:& \Phi_j=[\gamma + \varepsilon_j](\mbox{mod} 2\pi), \quad \gamma \in [0,2\pi),\\
	H_1&:& \Phi_j=[m(\Delta_j)+ \varepsilon_j](\mbox{mod} 2\pi),  \quad \exists \ j \ \mid \ m(\Delta_j)\neq\gamma+2l\pi  \  \forall \ l\in \mathbb{Z}.
\end{eqnarray*}
It will be assumed that the errors have zero mean and finite and constant concentration. In the linear response case, the test statistic was built by using the quadratic distance to measure the differences between the responses and the estimated curves under each of the hypotheses. In this case, it is not possible to use such distance since it is not well defined on the circle. Therefore, an appropriate circular distance must be employed. We propose to use the distance defined in (\ref{eq:circ_distance})  to construct the test statistic. As a result, the proposed statistic takes the form:
\[C_2=\frac{\sum_{j=1}^{n}[1-\cos(\Phi_j-\hat{\gamma})]-\sum_{j=1}^{n}[1-\cos(\Phi_j-\hat{m}(\Delta_j))]}{\sum_{j=1}^{n}[1-\cos(\Phi_j-\hat{m}(\Delta_j))]}.\]
where $\hat{\gamma}$ is the sample mean direction of the responses, given by
$$ \hat{\gamma}=\mbox{atan}2\left(\sum_{j=1}^{n}\sin \Phi_j,\sum_{j=1}^{n}\cos \Phi_j\right)$$
and $\hat{m}$ is the nonparametric estimator for circular responses~(\ref{eq:circ_mestimator}). 

The distribution of $C_2$ under $H_0$ is approximated through bootstrap methods. The resampling strategy is specified hereafter. (i) Given a smoothing parameter $h$ or $\kappa$ (depending on the nature of the predictor variable), compute the value of the statistic $C_2$ for the data, denoted by $Obs$. (ii) From the computed values of $\hat{\gamma}$, obtain the residuals under the null hypothesis ($\hat{\varepsilon}_{j}=\Phi_{j}-\hat{\gamma}$, $j\in\{1,...,n\}$) and construct the resampled responses as $\Phi_{j}^{*}=[\hat{\gamma}+\hat{\varepsilon}_{j}^*](\mbox{mod }2\pi)$, where $\hat{\varepsilon}_{j}^*$ are obtained from sampling the residuals randomly with replacement. (iii) With the same smoothing parameter as in (i), compute the value of the test statistic for the bootstrap resample, $C_2^{*(b)}$. (iv) Repeat (ii) and (iii) $B$ times to obtain $C_2^{*(1)},...,C_2^{*(B)}$, and approximate the critical value as $\sum_{b=1}^{B} \bm{1}_{\{C_2^{*(b)}\geq Obs\}}/B$, where $\bm{1}_A$ denotes the indicator function of $A$. 

It should be noticed that, as in any nonparametric test (see \cite{Bowman_Azzalini1997}), the outcome may be influenced by the smoothing parameter. An optimal smoothing parameter in terms of estimation might not be suitable for hypotheses testing, because of the bias present in the estimation of $m$. In practice, it is recommended to run the test over a sequence of smoothing parameters in a reasonable range. In the simulation study carried out in Section~\ref{sec:simulation}, we analyze the performance of the tests with different smoothing parameters derived from a cross-validation bandwidth/concentration.			

\section{ANCOVA models for circular regression}
\label{sec:ancova}
In this section we introduce ANCOVA models for circular regression, and testing tools for equality and parallelism. First, we focus on the circular predictors and real-valued responses case, while the circular response scenarios are analyzed later. A categorical covariate inducing $I$ groups will be now introduced in the model, each one identified by subscript $i=1,\ldots,I$. The number of data in the $i$th group will be denoted by $n_i$.

\subsection{Circular covariate}
\label{sec:circular_covariate}
An ANCOVA regression model for the circular-linear regression scenario is formulated as
\[Y_{ij}=m_i(\Theta_{ij})+\varepsilon_{ij}, \quad i\in \{1,...,I\}, \ j \in \{1,...,n_i\},\]
where the $\varepsilon_{ij}$ are, first, assumed to follow a $N(0,\sigma^2)$ distribution. In the following, two tests will be proposed, one for equality and one for parallelism. \\

\subsubsection{Test of equality}
The equality of the curves is tested through the following hypotheses statement:
\begin{eqnarray*}
	H_0&:& Y_{ij}=m(\Theta_{ij})+\varepsilon_{ij},  \quad \forall \ i\in\{1,...,I\},\\
	H_1&:& Y_{ij}=m_i(\Theta_{ij})+\varepsilon_{ij}, \quad \exists \ i,k \in\{1,...,I\} \mid  m_i(\cdot)\neq m_k(\cdot). 
\end{eqnarray*}
The corresponding test statistic takes the form 
\[C_3=\frac{1}{\hat{\sigma}^2}\sum_{i=1}^{I}\sum_{j=1}^{n_i}[\hat{m}_i(\Theta_{ij})-\hat{m}(\Theta_{ij})]^2.\]
A key point to obtain the expression of $C_3$ is the estimation of the variance. The circular nature of the predictor must be considered, given that it plays an important role when computing $\hat{\sigma}^2$. We propose estimating the variance by using periodic pseudoresiduals, which are a modification of the pseudoresiduals defined by \cite{Gasseretal1986}. Let $Y_{i[j]}$, with $j\in\{1,...,n_i\}$, denote the value of $Y$ corresponding to $\Theta_{i[j]}$, where $\Theta_{i[j]}$ represents the $j\mbox{th}$ smallest value on the real line of the sample from $\Theta$ in group $i$ (given that an origin has been chosen). The new pseudoresiduals are defined as 
\[\tilde{\varepsilon}_{i[j]}=\frac{\Theta_{i[j+1]}-\Theta_{i[j]}}{\Theta_{i[j+1]}-\Theta_{i[j-1]}}Y_{i[j-1]} + \frac{\Theta_{i[j]}-\Theta_{i[j-1]}}{\Theta_{i[j+1]}-\Theta_{i[j-1]}}Y_{i[j+1]} - Y_{i[j]}, \]
with $ i\in\{1,...,I\}, \ j\in \{1,...,n_i\}$. Here, we have
$Y_{i[n_i+1]}=Y_{i[1]}$, $ Y_{i[n_i+2]}=Y_{i[2]}$ and $Y_{i[0]}=Y_{i[n_i]}$. 
The periodic pseudoresiduals can then be expressed as $\tilde{\varepsilon}_{i[j]}=a_{i[j]}Y_{i[j-1]} + b_{i[j+1]} Y_{i[j+1]} - Y_{i[j]}$, and thus, the variance in each group and the total variance are estimated, respectively, as
\[\hat{\sigma}_i^2=\frac{1}{n_i}\sum_{j=1}^{n_i}\frac{1}{c^2_{i[j]}}\tilde{\varepsilon}_{i[j]}^2 \mbox{ and } \hat{\sigma}^2=\frac{1}{n-I}\sum_{i=1}^{I}n_i\hat{\sigma}_i^2, \]
where $c_{i[j]}^2=a_{i[j]}^2+b_{i[j]}^2+1$, $i\in\{1,...,I\}$, $j \in \{1,...,n_i\}$.
After some calculations, this estimator can be written in matrix-vector notation as $\bm{Y'KY}=\bm{Y'P'PY}$, where $\bm{P}$ is a $n \times n$ block matrix.\\

\subsubsection{Test of parallelism}
For testing parallel regression curves, the following hypotheses are used:
\begin{eqnarray*}
	H_0&:& Y_{ij}=\gamma_i+m(\Theta_{ij})+\varepsilon_{ij}, \quad \gamma_1=0 , \gamma_i \in \mathbb{R}, \forall \ i \in \{1,...,I\},\\
	H_1&:& Y_{ij}=m_i(\Theta_{ij})+\varepsilon_{ij}, \quad  \exists \ i,k\in\{1,...,I\} \mid   m_i(\cdot)\neq m_k(\cdot)+\gamma \ \forall \ \gamma\in \mathbb{R}. 
\end{eqnarray*}	
The next statistic is used to test the differences between the models under each one of the two hypotheses:
\[C_4=\frac{1}{\hat{\sigma}^2}\sum_{i=1}^{I}\sum_{j=1}^{n_i}[\hat{\gamma}_i+\hat{m}(\Theta_{ij})-\hat{m}_i(\Theta_{ij})]^2. \]
For estimating the shift parameter $\gamma$, the model is written in vector-matrix notation: 
\begin{equation}
\bm{Y}=\bm{D\gamma}+\bm{m}+\bm{\varepsilon},
\label{eq:model_circlin_parallelism}
\end{equation}
where $\bm{D}$ is a known matrix consisting of 0s and 1s. Given a vector $\bm{\gamma}$, an estimate of the regression function can be constructed:
$$ \bm{\hat{m}}=\bm{S}(\bm{Y}-\bm{D\gamma}),$$
with $\bm{S}$ being a smoothing matrix constructed with the circular-linear regression method. Substituting this estimator in equation (\ref{eq:model_circlin_parallelism}) and applying the least squares method, an estimate of $\bm{\gamma}$ is derived:
$$ \hat{\bm{\gamma}}=[\bm{D'}(\bm{I}_n- \bm{S_1})'(\bm{I}_n- \bm{S_1})\bm{D}]^{-1}\bm{D}'(\bm{I}_n- \bm{S_1})'(\bm{I}_n- \bm{S_1})\bm{Y}=\bm{WY}, $$
where $\bm{S_1}$ is a preliminary smoothing matrix. After $\bm{\hat{\gamma}}$ is obtained, the regression function $m$ is estimated as 
$$\bm{\hat{m}}=\bm{S}(\bm{Y}-\bm{D\hat{\gamma}}). $$

However, for estimating the vector of parameters $\bm{\hat{\gamma}}$ it is necessary to choose a first smoothing parameter $\kappa_1$, independent of the one used to estimate the actual curves, and the selection of this parameter is not trivial. In the real-valued case,  \citet{Bowman_Azzalini1997} use $2R/n$, where $R$ is the range of the design points, as this choice restricts the smoothing to approximately eight neighboring observations when the data are equally spaced (if a normal kernel is used).

With the objective of finding an adequate automatic rule in the circular setting, we use a local smoothing parameter, which showed a good performance in practice.
Let $d_2(\cdot,\cdot)$ be defined as
$$ d_2(\Phi,\Theta)=\min\{|\Phi-\Theta|,2\pi-|\Phi-\Theta|\}, \quad \Phi,\Theta\in[0,2\pi), $$
i.e., $d_2$ is the geodesic distance. Our proposal consists in finding a preliminary vector of smoothing parameters, $\bm{h_1}$, containing one parameter for each observation, in which the parameter associated to observation $\Theta_{ij}$, $h_{1;ij}$, will be the distance to its $8$th nearest neighbor (considering distance $d_2$). Then, $\bm{h_1}$ is used to obtain a vector of smoothing parameters valid for the circular case  using the results in \cite{Gumbeletal1953}, which show that for large values of $\kappa$ the von Mises $vM(\mu,\kappa)$ converges in distribution to a $N(\mu,1/\sqrt{\kappa})$. Thus, if $h_{1;ij}$ is the preliminary smoothing parameter corresponding to $\Theta_{ij}$, the concentration parameter for this observation will be $\kappa_{1;ij}=1/h_{1;ij}^2$. \\

\subsubsection{Distribution of the statistics}
In order to obtain the distributions of $C_3$ and $C_4$ under $H_0$ we must note that their numerators can be expressed, respectively, as
$$ \bm{Y}'[\bm{S_d}-\bm{S}]'[\bm{S_d}-\bm{S}]\bm{Y} $$
and
$$ \bm{Y}'[\bm{DW}+\bm{S}(\bm{I}_n-\bm{DW})-\bm{S_d}]'[\bm{DW}+\bm{S}(\bm{I}_n-\bm{DW})-\bm{S_d}]\bm{Y},$$
where $\bm{S}$ is the smoothing matrix under the equality (or parallelism) assumption and $\bm{S_d}$ is the block matrix constructed with the smoothing matrices for each group. Then, both statistics can be expressed in the form $ \bm{Y'QY}/\bm{Y'GY}$, where $\bm{Q}$ is a symmetric matrix and $\bm{G}$ is obtained straightforward from the variance estimator. Now, under the null hypothesis, if the same concentration parameter is used for obtaining the matrices $\bm{S}$ and $\bm{S_d}$, by analyzing the bias terms of the circular-linear estimator ($\hat{\beta}_0$ in (\ref{eq:nonpar_fit})), it can be shown that the distribution of $ \bm{Y'QY}/\bm{Y'GY}$ is almost equivalent to the distribution of $ \bm{\varepsilon'Q\varepsilon}/\bm{\varepsilon'G\varepsilon}$. Then, using the first three cumulants of $\bm{\varepsilon}'(\bm{Q}-\bm{G} \cdot Obs)\bm{\varepsilon}$ (where $Obs$ is the observed value of $C_3$ or $C_4$), the shifted and scaled $\chi^2$ approximation described in Section~\ref{subsec:noeffect} can be employed. 

Note that conditions over the residuals can be relaxed, assuming only zero mean and constant variance. In such case, the distribution of the statistics can be obtained through bootstrap methods, in a similar way as in the following scenarios.

\subsection{Circular response}
\label{sec:circular_response}
An ANCOVA model for a circular response variable can be expressed as 
\[
\Phi_{ij}=[m_i(\Delta_{ij})+\varepsilon_{ij}](\mbox{mod}2\pi), \quad i\in\{1,...,I\}, \ j\in \{1,...,n_I\}.
\] 
It will be assumed that the errors $\varepsilon_{ij}$ have zero mean and finite and constant concentration $\kappa$. The tests of equality and parallelism are presented next. \\

\subsubsection{Test of equality}
The hypotheses stated for testing the equality of the curves are	
{\small		
	\begin{eqnarray*}
		&H_0:& \Phi_{ij}=[m(\Delta_{ij})+\varepsilon_{ij}](\mbox{mod}2\pi), \quad \forall \ i\in\{1,...,I\},  \\
		&H_1:& \Phi_{ij}=[m_i(\Delta_{ij})+\varepsilon_{ij}](\mbox{mod}2\pi), \quad \exists \ i,k\in \{1,...,I\} \mid  m_i(\cdot)\neq m_k(\cdot)+ 2l\pi \ \forall \ l\in\mathbb{Z}.  
\end{eqnarray*}}
\hspace{-0.15cm}In pursuance of constructing the test statistic, one needs to measure the differences between the estimated regression functions under each of the hypotheses. Since $\hat{m}$ and $\hat{m}_i$ have circular nature, the distance defined in (\ref{eq:circ_distance}) will be adequate to construct the test statistic. However, a second problem is encountered, as a dispersion measure is needed to normalize the statistic. Our proposal is to estimate the circular variance, a measure of dispersion in the circle (see \citet{Mardia_Jupp2000}). Consequently, our proposed test statistic takes the form
\[
C_5= \frac{1}{\bar{D}}\sum_{i=1}^{I}\sum_{j=1}^{n_i}\big[1-\cos(\hat{m}_i(\Delta_{ij})-\hat{m}(\Delta_{ij}))\big],
\]
where $\bar{D}$ is an estimator of the circular variance  defined as 
\[
\bar{D}=\frac{1}{n-I}\sum_{i=1}^{I}\sum_{j=1}^{n_i}[1-\cos(\Phi_{ij}-\hat{m}_i(\Delta_{ij}))].
\]

\subsubsection{Test of parallelism}
When considering a circular response, both for circular or linear covariates, the regression curves might be parallel in a way in which the shape of the regression function is the same for all groups except for an angular shift. This behavior can be tested with the following hypotheses statement:
{\small
	\begin{eqnarray*}
		&H_0:& \Phi_{ij}=[\gamma_i + m(\Delta_{ij})+\varepsilon_{ij}](\mbox{mod}2\pi), \quad \gamma_1=0,  \forall \ i\in\{1,...,I\}, \\
		&H_1:& \Phi_{ij}=[m_i(\Delta_{ij})+\varepsilon_{ij}](\mbox{mod}2\pi), \quad \exists \ i,k\in \{1,...,I\} \mid m_i(\cdot)\neq [m_k(\cdot)+\gamma](\mbox{mod}2\pi),
\end{eqnarray*}}
\hspace{-0.15cm}where $\gamma_i \in[0,2\pi)$. As before, the circular distance (\ref{eq:circ_distance}) is used for constructing the test statistic. 
\[
C_6= \frac{1}{\bar{D}}\sum_{i=1}^{I}\sum_{j=1}^{n_i}\big[1-\cos(\hat{\gamma}_i+\hat{m}(\Delta_{ij})-\hat{m}_i(\Delta_{ij}))\big].
\]
Again, the shift parameters must be estimated. As a first step, a first smoothing parameter needs to be selected to obtain the preliminary estimator of the global regression function, namely $\hat{m}^1$, to minimize the bias in the preliminary estimation of $m$ and therefore in the estimation of $\gamma_1,...,\gamma_I$. Although it is recommended to explore several parameters, an automatic rule was derived. When the predictors are linear ($\Delta=X$), the rule consists of using a vector of smoothing parameters in which each of them corresponds to one observation. Each parameter will be the distance to the $8$th nearest observation. On the other hand, in the case where the predictor is of a circular nature ($\Delta=\Theta$), the rule is the same as in the test of parallelism for circular-linear regression (Section~\ref{sec:circular_covariate}). Then, the parameters $\gamma_1,\ldots,\gamma_I$ are estimated by solving the minimization problem
\begin{eqnarray}	
&&\arg\min_{\gamma_1,..., \gamma_I} \sum_{i=1}^{I} \sum_{j=1}^{n_i}[1-\cos(\Phi_{ij}-\gamma_i -\hat{m}^1(\Delta_{ij}))] \label{eq:Minimization_problem}\\
&&  \mbox{s.t. } \quad \ \ 
\gamma_i \in [0,2\pi), \ \forall \ i\in \{1,...,I\}. \nonumber
\end{eqnarray}

\begin{prop}
	For $i=1,\ldots,I$, let $C_i$ and $S_i$ denote $$C_i=\sum_{j=1}^{n_i} \cos (\Phi_{ij} -\hat{m}^1(\Delta_{ij})) \ \mbox{and} \ S_i=\sum_{j=1}^{n_i} \sin (\Phi_{ij} -\hat{m}^1(\Delta_{ij})).$$  
	If  for all $i=1,\ldots,I$ we have $C_i\neq0$ or $S_i\neq0$, then $(\hat{\gamma}_1,\ldots,\hat{\gamma}_I)$, where $\hat{\gamma}_i=\mbox{atan}2(S_i,C_i)$, $i=1,\ldots,I$,  is a global minimum of minimization problem (\ref{eq:Minimization_problem}).
	\label{prop: minimization_gamma}
\end{prop}

The proof of Proposition \ref{prop: minimization_gamma} is given in Appendix~\ref{ap:proof}. Note that the estimators $\hat{\gamma}_1,...,\hat{\gamma}_I$ obtained will not be unbiased, due to the bias of the preliminary estimator $\hat{m}^1$. However, the bias is smaller as the sample size increases.

\subsubsection{Distribution of the statistics}
The distribution of $C_5$ and $C_6$ under $H_0$ is obtained with bootstrap methods.  The resampling strategy is described next. (i) Choose a smoothing parameter, for example the one selected by cross-validation, to obtain the estimators $\hat{m}$ and $\hat{m}_1,...,\hat{m}_I$. (i, for the test of parallelism) Choose also a preliminary smoothing parameter $h_1$ or $\kappa_1$ (depending on the nature of the explanatory variable) and obtain the nonparametric estimator $\hat{m}^1$ and the shift parameter estimator $\hat{\bm{\gamma}}$. (ii) Compute the observed value of statistic $C_5$ or $C_6$, namely $Obs$. (iii) Obtain the residuals under the null hypothesis ($\hat{\varepsilon}_{ij}$) and construct the resampled responses ($\Phi_{ij}^{*}$) from the bootstrap residuals, $\hat{\varepsilon}_{ij}^*$, obtained from sampling the residuals randomly with replacement.
{\small
	\begin{eqnarray*}
		&&\mbox{(Equality)} :  \hat{\varepsilon}_{ij}=\Phi_{ij}-\hat{m}(\Delta_{ij}) \mbox{ and } \Phi_{ij}^{*}=[\hat{m}(\Delta_{ij})+\hat{\varepsilon}_{ij}^*](\mbox{mod }2\pi). \\
		&&\mbox{(Parallelism)} :  \hat{\varepsilon}_{ij}=\Phi_{ij}-\hat{\gamma}_i - \hat{m}(\Delta_{ij})  \mbox{ and } \Phi_{ij}^{*}=[\hat{\gamma}_i + \hat{m}(\Delta_{ij})+\hat{\varepsilon}_{ij}^*](\mbox{mod }2\pi). 
\end{eqnarray*}}
\hspace{-0.15cm}(iv) Using the smoothing parameter employed in (i) for estimating $\hat{m}$, compute the value of the test statistic for the bootstrap resample, $C_5^{*(b)}$ or $C_6^{*(b)}$. (iv for the test of parallelism) Use also the preliminary smoothing parameter obtained in (i). (v) Repeat (iii) and (iv) $B$ times to approximate the critical value as $\sum_{b=1}^{B} \bm{1}_{\{C_5^{*(b)}\geq Obs\}}/B$ or $\sum_{b=1}^{B} \bm{1}_{\{C_6^{*(b)}\geq Obs\}}/B$.
\section{Simulation study}
\label{sec:simulation}
Finite sample performance of the tests, both in terms of calibration and power, is explored by simulation. The no-effect tests, for the different regression scenarios, are investigated first. Tests for equality and parallelism are also analyzed for all the scenarios.

\subsection{Results for the no-effect tests}

For the no-effect tests, we simulate data from different regression models depending on the scenario, with sample size $n\in\{50,100,250,400\}$. The regression models for the different scenarios are the following:
\begin{itemize}
	\item \textit{Circular-linear:}  $Y=\beta \sin \Theta \cos \Theta + \varepsilon$, $\quad \beta=0,.25,.5$.
	\item \textit{Linear-circular:} $ \Phi=[3\pi/8 + \beta \cos(3X) + \varepsilon](\mbox{mod }2\pi)$, $\beta=0,.5,1$. 
	\item  \textit{Circular-circular:} $ \Phi=[3\pi/4 + \beta \sin(2\Theta + 2\sin (\Theta +\pi/2)) + \varepsilon](\mbox{mod }2\pi)$, $\beta=0,.35,.5$. 
\end{itemize}
When using the first value of $\beta$ in each model, data are simulated under the null hypothesis of no-effect of the predictor over the response. With the other two values of $\beta$, the alternative hypothesis holds. For the linear response case, the errors are drawn from a normal distribution with zero mean and standard deviation $.25$, which enables calibration by a shifted and scaled $\chi^2$ distribution. Rescaled exponential errors with rate parameter $5$  are also used, and in this case calibration is done though a bootstrap procedure.  When the response is circular, the errors are drawn from von Mises distributions with mean direction zero. The concentration is $\kappa=2$ for the model with linear predictors and $\kappa=4$ for the model with circular covariate. For calculating the percentage of rejections, the number of samples is 500. For the bootstrap procedure, the number of bootstrap replicates is 500. 

As mentioned in Section~\ref{subsec:noeffect}, the outcome of the tests may be seriously influenced by the smoothing parameter. Here, we study the performance of the tests when the smoothing parameter is selected by cross-validation ($cv$) and when we use other parameters which undersmooth or oversmooth the regression estimators. Specifically, when the covariate is circular, we use $cv/8$ and $4cv$ as the parameters which respectively oversmooth and undersmooth the estimated curve. In the linear-circular case, since the kernel used is linear, we consider the parameter $4cv$ for an oversmoothed estimator and $cv/4$ for an undersmoothed curve.

Percentages of rejection of the tests for a significance level of $\alpha=.05$ are displayed in Table~\ref{table:noeffect} for different sample sizes.  Further simulation results for significance levels $\alpha=.10$ and $\alpha=.01$ are provided in Tables~\ref{table:noeffect_010} and \ref{table:noeffect_001}. In what follows we will refer to the results for $\alpha=.05$. Focusing on the calibration of the tests, the smoothing parameters obtained by cross-validation do not provide a well calibration of the test under the null hypothesis, given that percentages of rejection obtained with this parameter are around 10\% in all cases for the first value of $\beta$. However, when using the other values of the bandwidth, percentages of rejection are close to the significance level $\alpha=.05$, being just slightly conservative when considering $4cv$ for $n=50$ in the circular-linear context.

On the other hand, the performance of the tests under the alternative is shown when the second and third values of $\beta$ are considered. In such cases, percentages of rejection tend to one as the sample size increases. The best performance is obtained, in general, when considering $cv$ as the smoothing parameter. Focusing on the two cases where the test seems to be well calibrated, the best performance is obtained when using an undersmoothed estimator. In that case, in the studied scenarios, the percentage of rejections is above $.2$, when $n=50$, $.5$, when $n=100$ and above $.98$, when $n=250$.

With the objective of comparing the two calibration alternatives in the circular-linear test, in Table~\ref{table:noeffect_switched_errors}, the shifted and scaled $\chi^2$ test is also employed with errors generated by the exponential distribution and the bootstrap calibration of the test is used with the normal errors. In general, similar results are obtained with both calibration methods. The $\chi^2$ test seems to be well calibrated (for $4cv$ and $cv/8$), even when errors are generated from the exponential distribution. Regarding the power, a slightly better behavior is observed, in general, with the bootstrap calibrated test.  When the objective is having a more efficient test (in computational terms), if the errors are normally distributed, since both test provide very similar percentages of rejections, the $\chi^2$ is the recommended calibration alternative.

\begin{table}[htbp]
	\caption{Percentages of rejection (for $\alpha=.05$) obtained with the no-effect tests using different smoothing parameters. Results for the first value of $\beta$ show empirical size, whereas results for the other values of $\beta$ show empirical power.}
	\label{table:noeffect}
	\begin{tabular}{cccccccccc}
		\hline
		\multicolumn{10}{c}{ Circular-Linear regression. $\chi^2$ calibration. Normal errors}\\
		\hline
		&  \multicolumn{3}{c}{$4cv$}&  \multicolumn{3}{c}{$cv$}&  \multicolumn{3}{c}{$cv/8$}\\
		\cmidrule(l){1-1} \cmidrule(l){2-4} \cmidrule(l){5-7} \cmidrule(l){8-10}
		$n$ & $\beta=0$ & $\beta=.2$ & $\beta=.3$  & $\beta=0$ & $\beta=.2$ & $\beta=.3$ & $\beta=0$ & $\beta=.2$ & $\beta=.3$ \\
		50 & .016 & .222 & .806 & .080 & .440 & .952 & .044 & .166 & .606 \\
		100 & .040 & .564 & .998  & .078 & .736 & 1 & .053 & .286 & .976 \\
		250 & .062 & .984 & 1 & .094 & .998 & 1  & .052 & .844 & 1 \\  
		400 & .064 & 1 & 1 & .106 & 1 & 1 &  .060 & .988 & 1 \\
		\hline
		\multicolumn{10}{c}{ Circular-Linear regression. Bootstrap calibration. Exponential errors}\\
		\hline
		&  \multicolumn{3}{c}{$4cv$}&  \multicolumn{3}{c}{$cv$}&  \multicolumn{3}{c}{$cv/8$}\\
		\cmidrule(l){1-1} \cmidrule(l){2-4} \cmidrule(l){5-7} \cmidrule(l){8-10}
		$n$ & $\beta=0$ & $\beta=.2$ & $\beta=.3$  & $\beta=0$ & $\beta=.2$ & $\beta=.3$ & $\beta=0$ & $\beta=.2$ & $\beta=.3$ \\
		50  & .040 & .396 & .922 & .090 & .648 & .994 & .058 & .294 & .814 \\
		100 & .034 & .820 & 1  & .094 & .916 & 1    &  .046 & .566 & .998  \\ 
		250 & .054 & 1    & 1   & .094 & 1    & 1    &   .052 & .980 & 1  \\
		400 & .058 & 1    & 1     & .096 & 1    & 1    & .042 & 1 & 1   \\	
		\hline
		\multicolumn{10}{c}{ Linear-Circular regression. Von Mises errors}\\
		\hline
		&  \multicolumn{3}{c}{$cv/4$}&  \multicolumn{3}{c}{$cv$}&  \multicolumn{3}{c}{$4cv$}\\
		\cmidrule(l){1-1} \cmidrule(l){2-4} \cmidrule(l){5-7} \cmidrule(l){8-10}
		$n$ & $\beta=0$ & $\beta=.5$ & $\beta=1$  & $\beta=0$ & $\beta=.5$ & $\beta=1$ & $\beta=0$ & $\beta=.5$ & $\beta=1$ \\
		50 & .042 & .410 & .980 & .070 & .494 & .772 & .056 & .474 & .756 \\
		100 & .066 & .734 & 1 & .100 & .808 & .974 & .068 & .802 & .974 \\
		250 & .044 & .978 & 1 & .066 & .990 & 1  & .046 & .990 & 1 \\
		400 & .044 & 1 & 1 & .082 & 1 & 1 & .054 & 1 & 1 \\
		\hline
		\multicolumn{10}{c}{ Circular-Circular regression. Von Mises errors}\\
		\hline
		&  \multicolumn{3}{c}{$4cv$}&  \multicolumn{3}{c}{$cv$}&  \multicolumn{3}{c}{$cv/8$}\\
		\cmidrule(l){1-1} \cmidrule(l){2-4} \cmidrule(l){5-7} \cmidrule(l){8-10}
		$n$ & $\beta=0$ & $\beta=.35$ & $\beta=.5$ & $\beta=0$ & $\beta=.35$ & $\beta=.5$ & $\beta=0$ & $\beta=.35$ & $\beta=.5$  \\
		50 & .048 & .244 & .462 & .118 & .436 & .712 &  .068 & .298 & .480 \\ 
		100 & .038 & .598 & .928 & .088 & .768 & .980 & .038 & .488 & .834 \\ 
		250 & .048 & .986 & 1 & .088 & .998 & 1  &  .044 & .934 & 1  \\
		400 & .058 & 1 & 1  & .102 & 1 & 1 & .058 & .998 & 1 \\ 
		\hline
	\end{tabular}
\end{table}

\subsection{Results for the ANCOVA tests}
The performance of the equality and parallelism tests will be illustrated in this section. For that purpose, data will be simulated from different models depending on the regression scenario, where each model has two groups with sample sizes $(n_1,n_2)\in \{(50,50),(50,100),(100,100),(100,250),(250,250)\}$. The regression models are the following:
\begin{itemize}
	\item \textit{Circular-linear:}\\
	Group 1: $Y=\cos\Theta\sin \Theta +\varepsilon$,\\
	Group 2: $Y=\beta \cos\Theta\sin \Theta +\varepsilon, \quad \beta=1,1.5,1.75$
	
	\item  \textit{Linear-circular:}\\
	Group 1: $\Phi=[2\sin(4X-1)+\varepsilon](\mbox{mod }2\pi)$,\\
	Group 2: $\Phi=[\beta\sin(4X-1)+\varepsilon](\mbox{mod }2\pi)$, $\quad \beta=2,1.75,1.5$.
	
	\item \textit{Circular-circular:}\\
	Group 1: $\Phi=[2\sin(2\Theta)+\varepsilon](\mbox{mod }2\pi)$,\\
	Group 2: $\Phi=[\beta\sin(2\Theta)+\varepsilon](\mbox{mod }2\pi)$, $\quad \beta=2,2.5,3$.
	
\end{itemize}

For the test of parallelism the same models are used, but a shift is added to the responses in the second group. The value of the shift parameter is $.2$ in the circular-linear case and $\pi/8$ in the tests for circular responses. As before, when the first value of $\beta$ is used the data are drawn from the null hypothesis and the alternative is considered if any of the other two values of $\beta$ is used. In the circular-linear regression case the $\chi^2$ calibration is applied to the simulated data with normally distributed errors, with zero mean and standard deviation $.25$. Rescaled exponential errors with rate parameter $5$ are also used, calibrating the distribution of the tests with the bootstrap procedure. For the tests with circular responses the errors are simulated from a von Mises distribution with mean zero and concentration parameter $\kappa=6$ for the test with a real-valued covariate and $\kappa=4$ in the circular-circular case. The number of samples, as well as the number of bootstrap replicates, is fixed to 500.

Percentages of rejection for a significance level of $\alpha=.05$ are shown in Table~\ref{table:ANCOVA} for different samples sizes, although results for $\alpha=.10$ and $\alpha=.01$ can be found, respectively, in Tables~\ref{table:ANCOVA_010} and \ref{table:ANCOVA_001}. In this case, the smoothing parameter applied was the one obtained by cross-validation ($cv$), but we also explore the performance of the tests with bandwidths that either undersmooth or oversmooth the estimated regression curves (see Tables~\ref{table:ANCOVA_undersmoothed} and \ref{table:ANCOVA_oversmoothed}).

Regarding the calibration of the tests, percentages of rejection lie around 5\% in all scenarios when $\alpha=.05$ and the $cv$ smoothing parameter are considered. In what refers to the power of the tests, when the data are drawn from the alternative hypothesis, percentages of rejection are closer to 1 as the sample size increases. However, the bootstrap calibration of the circular-linear test applied to exponential errors obtains low percentages of rejection (between 10\% and 20\%) for $n_1=50$. In all the other studied scenarios, the percentage of rejections is above $.3$, when considering the case $n_1=n_2=50$, above $.6$, when $n_1=n_2=100$, and above $.97$, when $n_1=n_2=250$.

In Table~\ref{table:ANCOVA_errors_switched}, it can be observed that the aforementioned behavior of the bootstrap parallelism test for circular-linear regression is also obtained when generating errors from the normal distribution (the percentage of rejections for $n_1=50$ under $H_1$ is around 10\%). Thus, it seems that a worse behavior is obtained when employing the bootstrap calibration instead of the $\chi^2$ test and the errors follow a normal distribution. Regarding the calibration of the $\chi^2$ test when errors follow the exponential distribution, we obtained that percentages of rejection under $H_0$ are slightly higher than $\alpha=.05$ (around 7\% or 8\%). Therefore, due to the anticonservative behavior, it is recommended that the calibration by the shifted and scaled $\chi^2$ distribution is only used when the normality assumption holds. 

As for the percentages of rejection obtained with other values of the smoothing parameters, the results in Tables~\ref{table:ANCOVA_undersmoothed} and \ref{table:ANCOVA_oversmoothed} show that with undersmoothed estimated regression curves, the percentages of rejection lie around the nominal level $\alpha=.05$ under $H_0$, although in that case the power of the tests is lower than when using the smoothing parameter selected by cross-validation. On the other hand, when oversmoothing, the tests are not well calibrated under $H_0$, obtaining very large percentages of rejection (even surpassing 30\% of rejections in some cases). 

\begin{table}[htbp]
	\caption{Percentages of rejection (for $\alpha=.05$) obtained with the ANCOVA tests using the smoothing parameters obtained by cross-validation. Results for the first value of $\beta$ show empirical size, whereas results for the other values of $\beta$ show empirical power.}
	\label{table:ANCOVA}
	\begin{tabular}{cccccccc}
		\hline
		\multicolumn{8}{c}{ Circular-Linear regression. $\chi^2$ calibration. Normal errors}\\
		\hline
		& & \multicolumn{3}{c}{Equality}&  \multicolumn{3}{c}{Parallelism}\\
		\cmidrule(l){1-2} \cmidrule(l){3-5} \cmidrule(l){6-8} 
		$n_1$ & $n_2$  & $\beta=1$ & $\beta=1.5$  & $\beta=1.75$ & $\beta=1$ & $\beta=1.5$  & $\beta=1.75$  \\
		50  & 50  & .054 & .522 & .916 & .054 & .592 & .928 \\
		50  & 100 & .040 & .718 & .978 & .046 & .732 & .978 \\
		100 & 100 & .072 & .902 &    1 & .054 & .926 &    1 \\
		100 & 250 & .050 & .984 &    1 & .056 & .986 &    1 \\
		250 & 250 & .044 &    1 &    1 & .054 &    1 &    1 \\
		\hline
		\multicolumn{8}{c}{ Circular-Linear regression. Bootstrap calibration. Exponential errors}\\
		\hline
		& & \multicolumn{3}{c}{Equality}&  \multicolumn{3}{c}{Parallelism}\\
		\cmidrule(l){1-2} \cmidrule(l){3-5} \cmidrule(l){6-8} 
		$n_1$ & $n_2$  & $\beta=1$ & $\beta=1.5$  & $\beta=1.75$ & $\beta=1$ & $\beta=1.5$  & $\beta=1.75$  \\
		50  & 50  & .040 & .744 & .970 & .076 & .120 & .216  \\
		50  & 100 & .056 & .844 & .994 & .054 & .132 & .204 \\
		100 & 100 & .046 & .980 & 1		 & .054 & .570 & .780  \\
		100 & 250 & .058 & 1 & 1    & .062 & 1 & 1     \\
		250 & 250 & .052 & 1    & 1    & .058 & 1    & 1     \\	
		\hline
		\multicolumn{8}{c}{ Linear-Circular regression. Von Mises errors}\\
		\hline
		& & \multicolumn{3}{c}{Equality}&  \multicolumn{3}{c}{Parallelism}\\
		\cmidrule(l){1-2} \cmidrule(l){3-5} \cmidrule(l){6-8} 
		$n_1$ & $n_2$ & $\beta=2$ & $\beta=1.75$ & $\beta=1.5$  & $\beta=2$ & $\beta=1.75$ & $\beta=1.5$ \\
		50  & 50  & .052 &	.304 &	.898 & .044 & .382 & .938 \\
		50  & 100 & .066 &	.450 &	.976 & .062 & .520 & .988 \\
		100 & 100 & .050 &	.650 &	.998 & .058 & .668 &    1 \\
		100 & 250 & .046 &	.784 &	   1 & .048 & .830 &    1 \\
		250 & 250 & .046 &	.976 &     1 & .056 & .970 &    1 \\
		\hline
		\multicolumn{8}{c}{ Circular-Circular regression. Von Mises errors}\\
		\hline
		& & \multicolumn{3}{c}{Equality}&  \multicolumn{3}{c}{Parallelism}\\
		\cmidrule(l){1-2} \cmidrule(l){3-5} \cmidrule(l){6-8} 
		$n_1$ & $n_2$ & $\beta=2$ & $\beta=2.5$ & $\beta=3$  & $\beta=2$ & $\beta=2.5$ & $\beta=3$ \\
		50  & 50  & .060 &	.408 &	.966 & .056 & .384 & .968 \\
		50  & 100 &	.062 &	.498 &	.994 & .062 & .562 & .996 \\
		100 & 100 & .052 &	.834 &	   1 & .048 & .838 &    1 \\
		100 & 250 & .064 &	.942 &	   1 & .072 & .950 &    1 \\
		250 & 250 & .064 &	   1 &     1 & .034 & .998 &    1 \\
		\hline
	\end{tabular} 
\end{table}


\section{Real data examples}
\label{sec:data}

The datasets described in Section~\ref{sec:intro} are used to illustrate the new proposals. We start by applying the new methods to the flywheels data. Then, the tests for circular responses are applied to the sand hoppers dataset.

\subsection{Flywheel data}
\label{subsec:flywheel}

Consider the flywheels data, described in the introduction, where the angle of imbalance of 60 devices was analyzed. Four different kinds of metals were employed in the production process, with 15 flywheels corresponding to each type of metal. Although the sample size for each group is small, this example is just meant to illustrate the techniques previously proposed.  

\begin{figure}[t]
	\centering
	\subfloat{
		\label{fig:data_anderson_total_param}
		\includegraphics[width=0.48\textwidth]{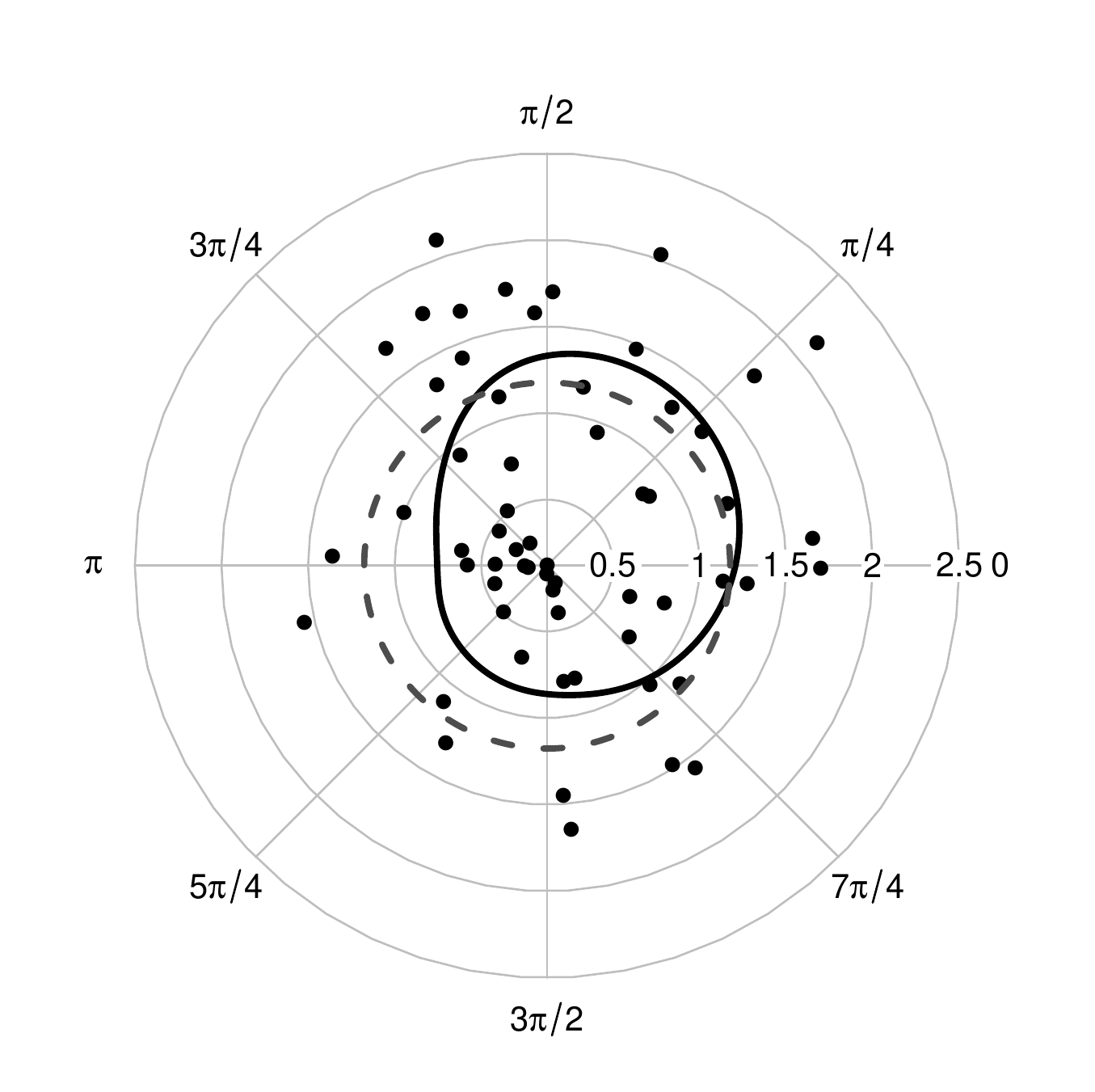}}
	\subfloat{
		\label{fig:data_anderson_groups_param}
		\includegraphics[width=0.48\textwidth]{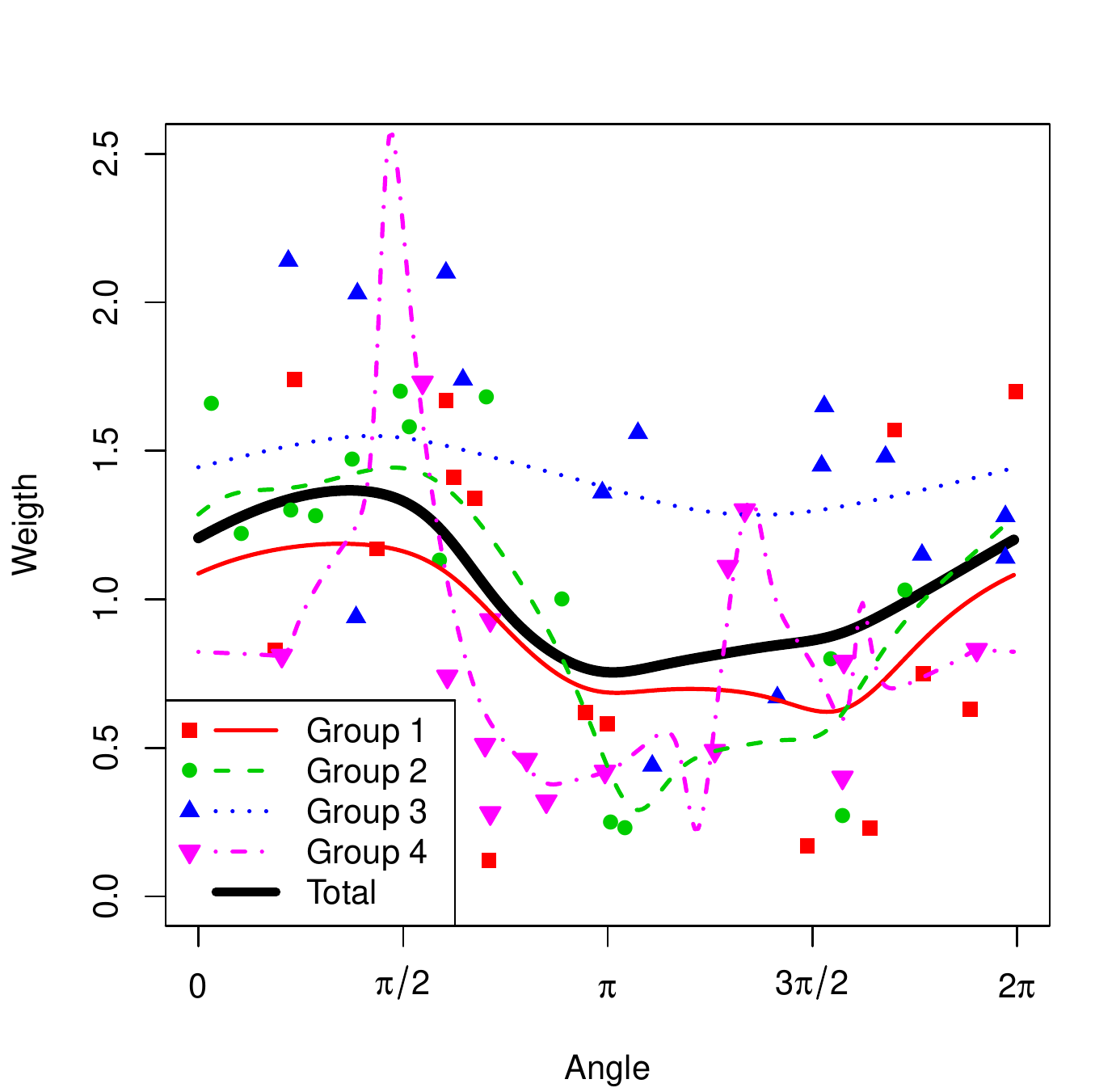}}
	\caption{Scatter plots of the angle of imbalance (in radians) against the balancing weight (in inch-ounces). Left: circular representation with estimated regression curve (continuous line) and estimated regression curve under the no-effect hypothesis (dashed line). Right: linear representation with estimated regression curve for the whole sample and for each group (as indicated in the legend). }
	\label{fig:data_anderson}
\end{figure}

A single nonparametric regression model can be constructed, without considering the different groups, as in the left panel of  Figure~\ref{fig:data_anderson} (continuous line), where the regression function was estimated with the circular-linear nonparametric estimator (see Section \ref{subsec:nonparametric}) using all the data. The dashed line represents the average of the responses, corresponding to the estimated model under the hypotheses of no-effect of the covariate.  The nonparametric estimation of the regression function changes for the different values of the predictor variable, but it could be possible that the responses did not depend on the angle of imbalance  and that the features of the curve were due to sample noise. To ascertain this,  the no-effect test for circular predictors (presented in Section~\ref{subsec:noeffect}) is applied to the data. The Kolmogorov-Smirnoff test was used to test the normality of the residuals, obtaining that the normality assumption was not rejected. Therefore, the $\chi^2$ approximation is used, but similar results are obtained with the bootstrap version of the test.  A range of smoothing parameters between 0 and 15 was considered, obtaining a $p$-value for each bandwidth. The results, showed in the top panel of Figure~\ref{fig:traces}, were lower than $0.05$ for all the concentration parameters considered, concluding, for this significance level, that the angle of imbalance is significant.

However, since the metal used in the molding is different, it could be possible to have different regression curves for the groups. The right panel in  Figure~\ref{fig:data_anderson} shows the data with different colors for each of the groups, with their corresponding estimated regression curves. The regression curve for all the data was represented in black. The smoothing parameter was selected by cross-validation, using only the data belonging to each group. 

The test of equality is first applied to the data, to test if all the regression curves are the same. Again, normality was not rejected for the residuals, so the $\chi^2$ calibration was applied. The value of the statistic obtained is $20.96$, while the $p$-value is  $.0263$, lower than the nominal level $\alpha=.05$. Thus, for that significance level, the hypothesis of equal regression curves is rejected. This result is obtained using the concentration parameter selected by cross-validation for all the data ($2.85$). For a better application of the test, a sequence of concentration values is used, obtaining, as shown in Figure~\ref{fig:traces} (top panel), that the equality assumption is not rejected for concentration values approximately larger than 5, although given the sample size, large smoothing parameters are quite unrealistic in practice. Then, it can be concluded that there is evidence for saying that the four regression curves are not equal for a significance level of $.05$.

Once the equality hypothesis is rejected, it could be checked if the regression curves are parallel. The parallelism test is applied with the smoothing parameter selected by cross-validation ($2.85$), and the obtained value for the test statistic is $5.44$, while the $p$-value is $0.4695$, much greater than $\alpha=.05$. Thus, there is no evidence for rejecting the null hypothesis of parallel regression curves. To avoid compromising results because of the selection of the smoothing parameter, the test is applied using a range of smoothing values. The trace of the test shows that the null hypothesis is not rejected for $\alpha=.05$ for any of the smoothing parameters considered ($\kappa$ lying between .05 and 15), as it can be seen in Figure~\ref{fig:traces} (top panel).

\subsection{Sand hoppers data}
\label{subsec:sandhoppers}

In the following, our goal is to apply the nonparametric significance test proposed in Section~\ref{subsec:noeffect} and the ANCOVA tests proposed in  Section~\ref{sec:circular_response} to the sand hoppers data. We will consider two different regression models, in both of which the response variable will be the direction of movement. The predictor variables will be the temperature and the sun azimuth. For the ANCOVA models, the type of arena will be the factor variable considered, which determines two groups: the unscreened and the screened sand hoppers. In our study we will only consider the male animals and the observations which took place in October. The total number of observations is 261, with 125 belonging to the unscreened group and 136 in the screened group.

\begin{figure}
	\centering
	\subfloat{
		\includegraphics[width=0.49\textwidth]{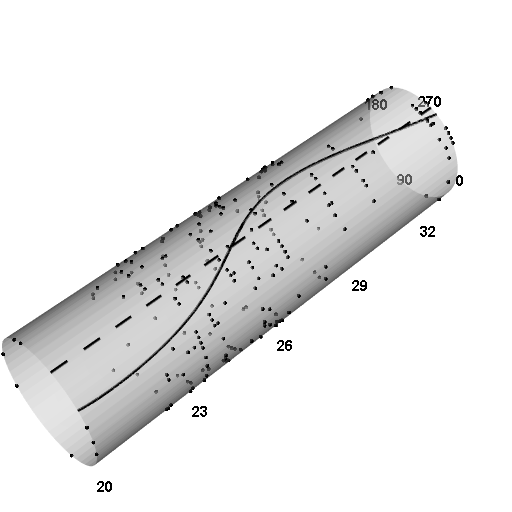}}
	\subfloat{
		\includegraphics[width=0.49\textwidth]{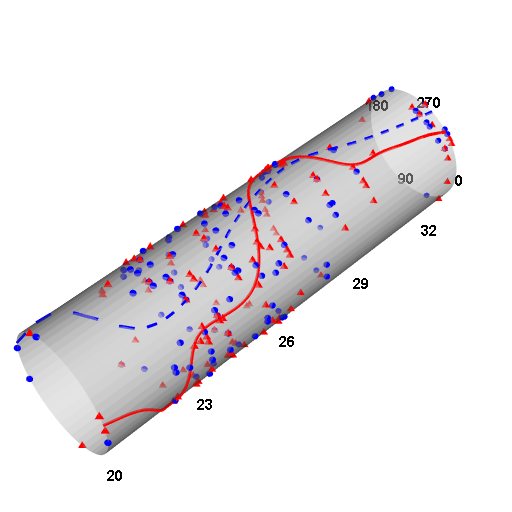}}
	\\
	\vspace{-0.5cm}
	\subfloat{
		\includegraphics[width=0.49\textwidth]{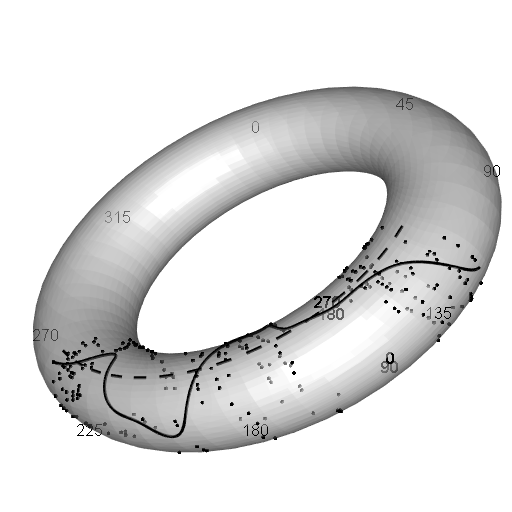}}
	\subfloat{
		\includegraphics[width=0.49\textwidth]{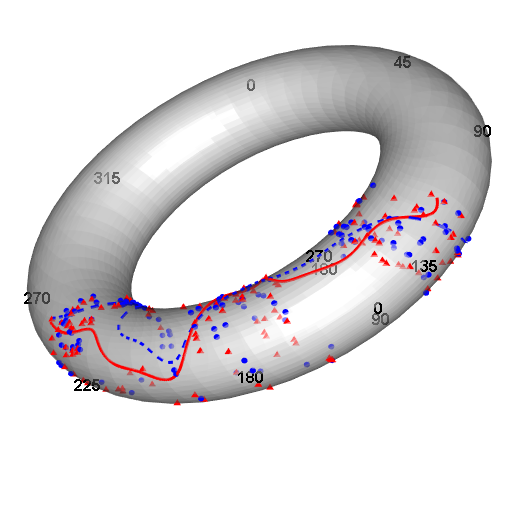}}
	\caption{Representations on the cylinder (top) and on the torus (bottom) of the sand hoppers data. Left column: estimated regression curve (continuous line)  and estimated regression curve under the no-effect hypothesis (dashed line) for the whole sample. Right column: scatter plot (points) and estimated regression curves for each group; screened group with triangles and continuous line (red in the colour version); unscreened group with circles and dashed line (blue in the colour version).}
	\label{fig:data_sandhoppers}
\end{figure}

To begin with, the relationship between the angle of direction of the sand hoppers and the temperature will be analyzed. The top-left panel in  Figure~\ref{fig:data_sandhoppers} shows a representation on the cylinder of the angle of direction against temperature, with the estimated regression curve obtained with the cross-validation method for selecting the bandwidth. The no-effect curve, i.e. a curve representing the global mean direction of the responses, is also represented. The first goal here is to ascertain if the temperature actually has an effect on the responses, for which the no-effect test for linear-circular regression is used. The test was applied using 1000 bootstrap replicates and over a sequence of smoothing parameters between $.01$ and $50$. The $p$-value was smaller than $\alpha=.05$ for all the smoothing parameters $h<9$, being the bandwidth value obtained by cross-validation equal to $2.98$. Therefore, we have evidences to state that the temperature affects the preferred direction of the sand hoppers.

Once it is known that the direction of movement is actually influenced by the temperature, the question relies on whether the regression functions for the screened and the unscreened animals are the same.  The top-right panel in Figure~\ref{fig:data_sandhoppers} shows representations of the data distinguishing between the screened and the unscreened groups, with the estimated regression functions. The smoothing parameter was selected by cross-validation in each group.

\begin{figure}
	\centering
	\includegraphics[width=.88\textwidth]{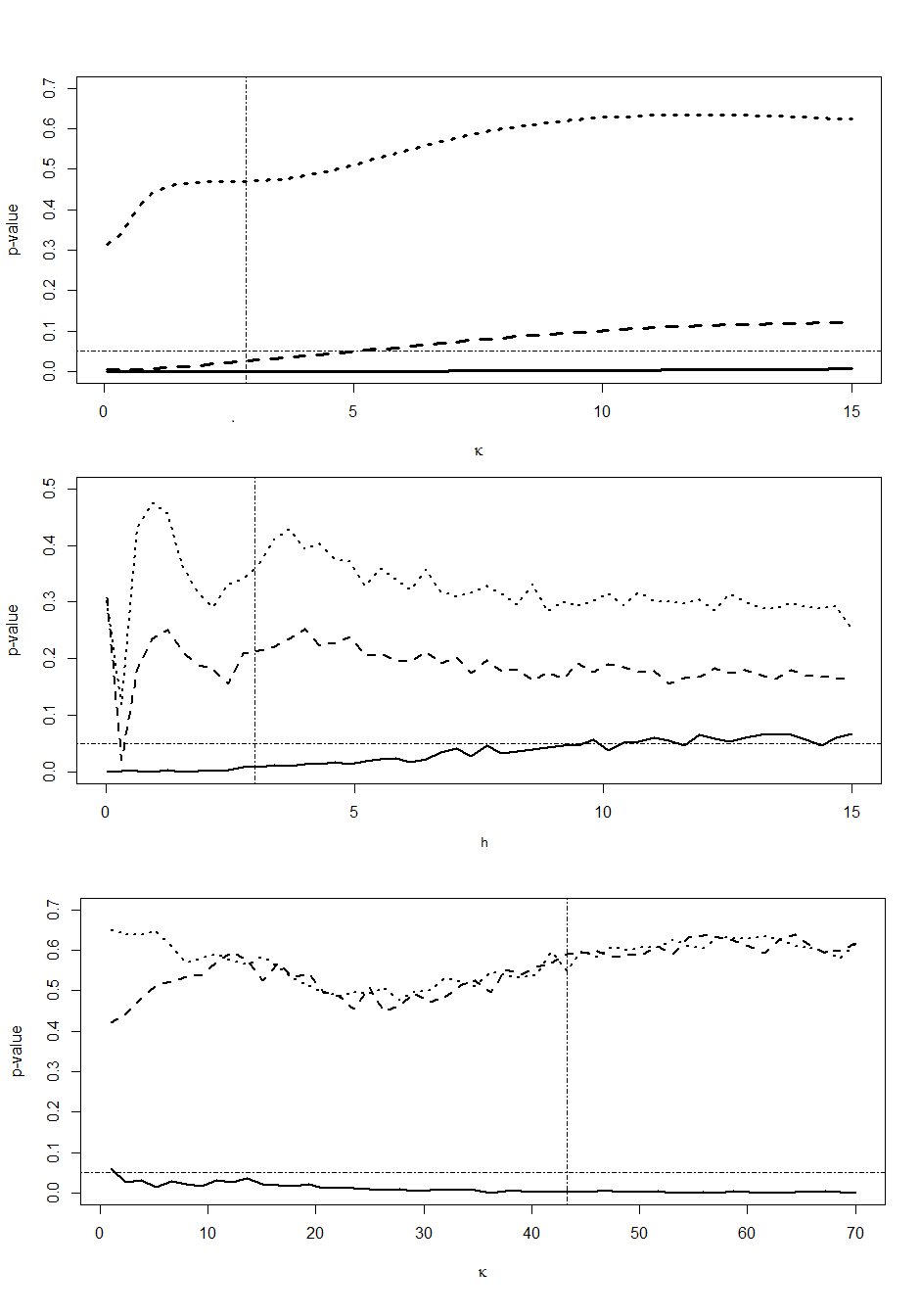}
	\vspace{-0.5cm}
	\caption{Traces of the no-effect tests (continuous line) equality tests (dashed line) and parallelism tests (dotted line) for the flywheels data (top) and sand hoppers data when the regressor variable is temperature (middle) and sun azimuth (bottom). Horizontal dashed-dotted line represents the significance level $\alpha=.05$ and vertical dashed-dotted line indicates the smoothing parameter selected by cross-validation in each scenario. }
	\label{fig:traces}
\end{figure}

The plots suggest that the behavior of the screened animals could be different from the behavior of the unscreened sand hoppers. This issue can be assessed by using the nonparametric test of equality for linear-circular regression. The test was applied to the data with the smoothing parameter selected by cross-validation ($2.98$) and using 1000 bootstrap replicates, obtaining a critical value of $.234$. Then, there are not evidences for a significance value $\alpha=.05$ to conclude that the two regression curves are different.

As mentioned before, our conclusion is that there are no evidences to reject that both curves are equal. For illustrative purposes, we can also see that there are no evidences against the hypothesis of parallelism by applying our proposed test, in which we obtain a $p$-value of $.357$ when using the smoothing parameter selected by cross-validation and 1000 bootstrap replicates.  As it was already stated, it is recommended to run the test over a sequence of smoothing parameters obtaining the trace of the tests, which are shown in the middle panel of Figure~\ref{fig:traces} for a sequence of 50 bandwidths ranging from $.01$ to $15$. The corresponding $p$-values for the tests of equality and parallelism were higher than $\alpha=.05$ for all the parameters considered (except for one in the equality test).

Now the regression relationship between the direction of movement and the sun azimuth will be studied. The bottom-left panel in Figure~\ref{fig:data_sandhoppers} displays a representation of the direction of movement against the sun azimuth on the torus, with the estimation of the regression function using cross-validation to select the smoothing parameter. The first objective is to determine if the sun azimuth affects the escape direction of the animals. For such purpose it is necessary to consider several concentration parameters in order to apply the no-effect test for circular-circular regression. A number of 1000 replicates was used for the bootstrap procedure, obtaining $p$-values lower than $.05$ for all the  considered values of the smoothing parameter (ranging from 1 to 70), being the cross-validation smoothing parameter $43.26$, as showed in the bottom panel of Figure~\ref{fig:traces}. Therefore, for that significance level it is rejected that the sun azimuth has no effect on the direction of movement of the sand hoppers.

The next objective consists on studying if the relationship between the direction of movement and the sun azimuth is different for the two groups of sand hoppers. The estimated regression curves for each group are represented in the bottom-right panel of Figure~\ref{fig:data_sandhoppers}, where the smoothing parameter was selected by applying the cross-validation method in each group.  When the cross-validation parameter for all data ($43.26$) is used, the $p$-value of the equality test is $.576$, much higher than the significance level $\alpha=.05$, concluding for this value of $\alpha$ that the regression curves are not significantly different. When applying the parallelism test (just for illustration), the obtained $p$-value was $.572$, concluding for $\alpha=.05$ that the regression curves are not significantly different. Figure~\ref{fig:traces} shows the traces of the equality and parallelism tests. For a significance level of $\alpha=.05$, the tests of equality and parallelism are not rejected for any of the considered concentration parameters (ranging from 1 to 70). \\


\section{Discussion}
\label{sec:conclusions}
This paper has been focused on different hypotheses testing problems for regression involving circular variables. In addition to surveying the existing nonparametric (kernel) regression models for this kind of data, new proposals for significance tests and ANCOVA tests have been introduced. Following some of the ideas in \cite{Bowman_Azzalini1997} and \cite{Young_Bowman1995}, we proposed different tests for no-effect and ANCOVA models. It should be noted that the circular nature of the response or the covariate leads important changes in the construction of the different tests. Throughout this paper we have overcomed the difficulties that arise from this non-euclidean support. As showed in the simulation study, a satisfactory performance has been assessed with our proposals. The importance of considering the test adapted to the circular regression context has been also showed in the illustration of the real datasets in the mechanical and ecological field. \\

As it has been mentioned along the paper, the smoothing parameter (bandwidth or concentration, depending on the type of the kernel) may present a relevant impact on the results of the tests. That is why we recommend to explore a range of bandwidths (as done for the illustration with real data). However, the use of a cross-validation bandwidth, obtained for estimation purposes in the different contexts, usually yields a reasonable performance of the ANCOVA tests.\\

\texttt{R} functions have been programmed for all the proposed methods, using previously programmed functions for the estimation of the regression curves in the circular context available in package \texttt{NPCirc} \citep{Oliveira_etal2014}. The code is currently available from the authors under request.\\

Regarding possible extensions of these tests to regression models involving more than a single real and/or circular explanatory variables, also including a categorical covariate, are also feasible. For example, a model studying the direction of movement with both the sun azimuth and the temperature in the sand hoppers dataset would be of interest. The suitable adaptations would include the use of different types of linear/circular weights, which could be considered as product kernels in the nonparametric estimators \citep{DiMarzioetal2009}. However, one should be aware that suitable smoothing parameters must be chosen in the new scenarios and although cross-validation ideas could be applied, the increasing dimension makes a thorough analysis more complex.\\

Finally, spherical regression models may be also considered. Similar ideas used in this paper to construct no-effect and equality and parallelism tests could be adapted to handle spherical responses and/or covariates. As a key tool for deriving the corresponding tests statistics, the nonparametric regression estimators introduced by \cite{DiMarzioetal2014} could be employed.

\section{Acknowledgements}
\label{sec:acknow}
The authors acknowledge the financial support of Project MTM2016--76969--P from the AEI co--funded by the European Regional Development Fund (ERDF), the Competitive Reference Groups 2017--2020 (ED431C 2017/38) from the Xunta de Galicia through the ERDF. Research of Mar\'ia Alonso-Pena was partially funded by the Institute of Mathematics (University of Santiago de Compostela). Jose Ameijeiras-Alonso was supported by the FWO research project G.0826.15N (Flemish Science Foundation) and GOA/12/014 project (Research Fund KU Leuven). We also thank Prof. Felicita Scapini for providing the sand hoppers data (collected under the support of the European Project  ERB ICI8-CT98-0270) and the Supercomputing Center of Galicia (CESGA) for the computational facilities. 






\appendix

\section{Proof of Proposition 3.1}\label{ap:proof}

The minimization problem in equation (\ref{eq:Minimization_problem}) is equivalent to the next maximization problem:

\begin{eqnarray*}	
	&&\arg\max_{\gamma_1,..., \gamma_I} \sum_{i=1}^{I} \sum_{j=1}^{n_i}\cos(\Phi_{ij}-\gamma_i -\hat{m}^1(\Delta_{ij})) \\
	&&  \mbox{s.t. } \quad \ \ 
	\gamma_i \in [0,2\pi), \ \forall \ i\in \{1,...,I\}.
\end{eqnarray*}
The objective function can be written as
\begin{eqnarray*}
	&& \Psi(\gamma_1,...,\gamma_I)=\sum_{i=1}^{I} \sum_{j=1}^{n_i}\cos(\Phi_{ij} -\hat{m}^1(\Delta_{ij})-\gamma_i)  \\
	&& = \sum_{i=1}^{I} \cos \gamma_i \sum_{j=1}^{n_i} \cos (\Phi_{ij} -\hat{m}^1(\Delta_{ij}))  + \sum_{i=1}^{I} \sin \gamma_i  \sum_{j=1}^{n_i} \sin (\Phi_{ij} -\hat{m}^1(\Delta_{ij})). 
\end{eqnarray*}
The partial derivative of $\Psi(\gamma_1,...,\gamma_I)$ with respect to $\gamma_t$, $t\in \{1,...,I\}$, is

\begin{eqnarray*}	
	\frac{\partial}{\partial \gamma_t}\Psi(\gamma_1,...,\gamma_I)  = - \sin \gamma_t  \sum_{j=1}^{n_t} \cos (\Phi_{tj} -\hat{m}^1(\Delta_{tj})) +  \cos \gamma_t  \sum_{j=1}^{n_t} \sin (\Phi_{tj} -\hat{m}^1(\Delta_{tj})). 
\end{eqnarray*}
By denoting
$$C_t=\sum_{j=1}^{n_t} \cos (\Phi_{tj} -\hat{m}^1(\Delta_{tj}))   \quad \mbox{and} \quad S_t=\sum_{j=1}^{n_t} \sin (\Phi_{tj} -\hat{m}^1(\Delta_{tj})),  $$
we have
\begin{equation}
\frac{\partial}{\partial \gamma_t}\Psi(\gamma_1,...,\gamma_I)  =0 \iff \sin \gamma_t C_t = \cos \gamma_t S_t.
\label{eq:proof_iff}
\end{equation}
Therefore, if $C_t\neq0$ or $S_t\neq 0$ we obtain 
$$ \hat{\gamma}_t= \mbox{atan}2(S_t,C_t). $$
In the case where $C_t=S_t=0$, we have $\frac{\partial}{\partial \gamma_t}\Psi(\gamma_1,...,\gamma_I)  =0 \ \forall \ (\gamma_1,\gamma_2,\ldots,\gamma_I)\in [0,2\pi)^I$. Consequently, if for all $i=1,\ldots,I$,  $C_i\neq0$ or $S_i\neq 0$, beacuse of equation (\ref{eq:proof_iff}) and Fermat's Theorem we have that  $(\hat{\gamma}_1,\hat{\gamma}_2,...,\hat{\gamma}_I)$ is the only critical point of $\Psi$.

To asses that $(\hat{\gamma}_1,\hat{\gamma}_2,...,\hat{\gamma}_I)$ is a maximum, we obtain the Hessian matrix of $\Psi$. If $l\in \{1,...,I\}$ and $l\neq t$, we have
$$ \frac{\partial^2}{\partial \gamma_t \gamma_l}\Psi(\gamma_1,...,\gamma_I)=0. $$
On the other side,
$$ \frac{\partial^2}{\partial \gamma_t^2}\Psi(\gamma_1,...,\gamma_I)=-\cos \gamma_t C_t - \sin \gamma_t S_t, $$
and
\begin{eqnarray*}
	&&\frac{\partial^2}{\partial \gamma_t^2}\Psi(\gamma_1,...,\gamma_I)\bigg |_{(\hat{\gamma}_1,\hat{\gamma}_2,...,\hat{\gamma}_I)}=-\cos \hat{\gamma}_t C_t - \sin \hat{\gamma}_t S_t  \\
	&&= - \frac{C_t}{\sqrt{S_t^2+C_t^2}}  C_t - \frac{S_t}{\sqrt{S_t^2+C_t^2}} S_t = \frac{-(S_t^2+C_t^2)}{\sqrt{S_t^2+C_t^2}}.
\end{eqnarray*}\\

As a result, the Hessian matrix at $(\hat{\gamma}_1,\hat{\gamma}_2,...,\hat{\gamma}_I)$ is given by

\[\bm{H}(\hat{\gamma}_1,\hat{\gamma}_2,...,\hat{\gamma}_I)=
\begin{pmatrix}
\frac{-(C_1^2+S_1^2)}{\sqrt{S_1^2+C_1^2}} & 0 & ... & 0\\
0 & \frac{-(C_2^2+S_2^2)}{\sqrt{S_2^2+C_2^2}} & ... & 0 \\
\vdots & \vdots & & \vdots \\
0 & 0 & ...&  \frac{-(C_I^2+S_I^2)}{\sqrt{S_I^2+C_I^2}}
\end{pmatrix}.\] \\

Then, $(\hat{\gamma}_1,\hat{\gamma}_2,...,\hat{\gamma}_I)$ is a maximum of $\Psi$ as all the eigenvalues of $\bm{H}(\hat{\gamma}_1,\hat{\gamma}_2,...,\hat{\gamma}_I)$ are negative. Since we obtained that if for all $i=1,\ldots,I$ $C_i\neq0$ or $S_i\neq 0$ the vector $(\hat{\gamma}_1,\hat{\gamma}_2,...,\hat{\gamma}_I)$ was the only critical point, we have that under that hypothesis it is the global maximum.

\section{Supplementary tables for simulation results}\label{ap:simus}

The contents of this section concern the simulation study conducted in Section 4. We provide simulation results regarding the models analyzed for the no-effect tests (Section 4.1) and ANCOVA tests (Section 4.2). Tables~\ref{table:noeffect_010} and \ref{table:noeffect_001} contain percentages of rejection of the no-effect tests with a significance level of $\alpha=.10$ and $\alpha=.01$, respectively (results for $\alpha=.05$ are provided in the main text). In addition, Tables~\ref{table:ANCOVA_010} and \ref{table:ANCOVA_001} show percentages of rejection for the ANCOVA test for the same significance levels. 

Additionally, for the \emph{circular-linear} regression case, results for the no-effect and ANCOVA tests are obtained considering a shifted and scaled $\chi^2$ distribution for calibration when using normal errors and results for those tests calibrated by the bootstrap approach when using exponential errors. We present percentages of rejection for the same tests when switching the distribution of the errors: Table \ref{table:noeffect_switched_errors} presents results for the no-effect test calibrated with the $\chi^2$ distribution applied to exponential errors and the test calibrated by bootstrap applied to normal errors. In the same line, Table~\ref{table:ANCOVA_errors_switched} collects percentages of rejection for the ANCOVA tests calibrated with the $\chi^2$ distribution applied to exponential errors and the bootstrap version of the tests applied to normal errors.

To conclude, we also study the performance of the ANCOVA tests when using values of the smoothing parameter different from the ones obtained by cross-validation. Table~\ref{table:ANCOVA_undersmoothed} contains percentages of rejection for the ANCOVA tests when using smoothing parameters which undermooth the regression curves. The finite sample performance of the tests when oversmoothing the regression curves is displayed in Table~\ref{table:ANCOVA_oversmoothed}.
\clearpage

\begin{table}[htbp]
	\begin{center}
		\caption{Percentages of rejection (for $\alpha=.10$) obtained with the no-effect tests using different smoothing parameters. Results for the first value of $\beta$ show empirical size, whereas results for the other values of $\beta$ show empirical power.}
		\label{table:noeffect_010}
		\resizebox{\textwidth}{!}{
			\footnotesize
			\begin{tabular}{cccccccccc}
				\hline
				\multicolumn{10}{c}{ Circular-Linear regression. $\chi^2$ calibration. Normal errors}\\
				\hline
				&  \multicolumn{3}{c}{$4cv$}&  \multicolumn{3}{c}{$cv$}&  \multicolumn{3}{c}{$cv/8$}\\
				\cmidrule(l){1-1} \cmidrule(l){2-4} \cmidrule(l){5-7} \cmidrule(l){8-10}
				$n$ & $\beta=0$ & $\beta=.2$ & $\beta=.3$  & $\beta=0$ & $\beta=.2$ & $\beta=.3$ & $\beta=0$ & $\beta=.2$ & $\beta=.3$ \\
				50 & .074 & .384 & .898  & .166 & .584 & .980 & .100 & .280 & .736 \\ 
				100 & .084 & .726 & 1  & .174 & .852 & 1 & .108 & .430 & .990 \\
				250 & .132 & .998 & 1  & .200 & 1 & 1 &  .114 & .920 & 1 \\
				400 & .128 & 1 & 1  & .204 & 1 & 1 & .124 & .996 & 1 \\ 
				\hline
				\multicolumn{10}{c}{ Circular-Linear regression. Bootstrap calibration. Exponential errors}\\
				\hline
				&  \multicolumn{3}{c}{$4cv$}&  \multicolumn{3}{c}{$cv$}&  \multicolumn{3}{c}{$cv/8$}\\
				\cmidrule(l){1-1} \cmidrule(l){2-4} \cmidrule(l){5-7} \cmidrule(l){8-10}
				$n$ & $\beta=0$ & $\beta=.2$ & $\beta=.3$  & $\beta=0$ & $\beta=.2$ & $\beta=.3$ & $\beta=0$ & $\beta=.2$ & $\beta=.3$ \\
				50  & .110 & .406 & .902 & .164 & .762 & .994 & .090 & .584 & .972  \\
				100 & .098 & .712 & .998 & .182 & .956 & 1 & .098 & .896 & 1 \\
				250 &  .126 & .992 & 1 & .188 & 1 & 1 & .118 & 1 & 1 \\ 
				400 & .120 & 1 & 1  & .188 & 1 &  1 & .128 & 1 & 1 \\
				\hline
				\multicolumn{10}{c}{ Linear-Circular regression. Von Mises errors}\\
				\hline
				&  \multicolumn{3}{c}{$cv/4$}&  \multicolumn{3}{c}{$cv$}&  \multicolumn{3}{c}{$4cv$}\\
				\cmidrule(l){1-1} \cmidrule(l){2-4} \cmidrule(l){5-7} \cmidrule(l){8-10}
				$n$ & $\beta=0$ & $\beta=.5$ & $\beta=1$  & $\beta=0$ & $\beta=.5$ & $\beta=1$ & $\beta=0$ & $\beta=.5$ & $\beta=1$ \\
				50 & .090 & .522 & .996 & .142 & .618 & .868 & .108 & .590 & .856 \\
				100 & .124 & .836 & 1 & .174 & .894 & .990 & .126 & .882 & .986 \\
				250 & .110 & .990 & 1 & .146 & .996 & 1 & .100 & .996 & 1 \\
				400 & .110 & 1 & 1 & .164 & 1 & 1 & .122 & 1 & 1 \\
				\hline
				\multicolumn{10}{c}{ Circular-Circular regression. Von Mises errors}\\
				\hline
				&  \multicolumn{3}{c}{$4cv$}&  \multicolumn{3}{c}{$cv$}&  \multicolumn{3}{c}{$cv/8$}\\
				\cmidrule(l){1-1} \cmidrule(l){2-4} \cmidrule(l){5-7} \cmidrule(l){8-10}
				$n$ & $\beta=0$ & $\beta=.35$ & $\beta=.5$ & $\beta=0$ & $\beta=.35$ & $\beta=.5$ & $\beta=0$ & $\beta=.35$ & $\beta=.5$  \\
				50 &  .102 & .408 & .670  & .216 & .598 & .828 & .144 & .424 & .614  \\
				100 & .094 & .746 & .976 & .158 & .874 & .994 & .118 & .626 & .910 \\
				250 & .088 & .996 & 1 & .168 & 1 & 1 & .110 & .976 & 1  \\
				400 & .138 & 1 & 1 & .210 & 1 & 1 & .122 & 1 & 1 \\ 
				\hline
		\end{tabular}}
		
	\end{center}
\end{table}

\begin{table}[htbp]
	\begin{center}
		\caption{Percentages of rejection (for $\alpha=.01$) obtained with the no-effect tests using different smoothing parameters. Results for the first value of $\beta$ show empirical size, whereas results for the other values of $\beta$ show empirical power.}
		\label{table:noeffect_001}
		\resizebox{\textwidth}{!}{
			\footnotesize
			\begin{tabular}{cccccccccc}
				\hline
				\multicolumn{10}{c}{ Circular-Linear regression. $\chi^2$ calibration. Normal errors}\\
				\hline
				&  \multicolumn{3}{c}{$4cv$}&  \multicolumn{3}{c}{$cv$}&  \multicolumn{3}{c}{$cv/8$}\\
				\cmidrule(l){1-1} \cmidrule(l){2-4} \cmidrule(l){5-7} \cmidrule(l){8-10}
				$n$ & $\beta=0$ & $\beta=.2$ & $\beta=.3$  & $\beta=0$ & $\beta=.2$ & $\beta=.3$ & $\beta=0$ & $\beta=.2$ & $\beta=.3$ \\
				50 &  0 & .018 & .832  & .008 & .196 & .832 & .006 & .036 & .370 \\
				100 &  .004 & .296 & .998 & .020 & .488 & .998 & .010 & .134 & .872 \\
				250 & .008 & .928 & 1 & .020 & .978 & 1 & .012 & .648 & 1 \\
				400 & .010 & .998 & 1  & .028 & 1 & 1 & .012 & .898 & 1  \\
				\hline
				\multicolumn{10}{c}{ Circular-Linear regression. Bootstrap calibration. Exponential errors}\\
				\hline
				&  \multicolumn{3}{c}{$4cv$}&  \multicolumn{3}{c}{$cv$}&  \multicolumn{3}{c}{$cv/8$}\\
				\cmidrule(l){1-1} \cmidrule(l){2-4} \cmidrule(l){5-7} \cmidrule(l){8-10}
				$n$ & $\beta=0$ & $\beta=.2$ & $\beta=.3$  & $\beta=0$ & $\beta=.2$ & $\beta=.3$ & $\beta=0$ & $\beta=.2$ & $\beta=.3$ \\
				50  &.010 & .100 & .582  &  .020 & .370 & .930 &  .006 & .162 & .774   \\
				100 &  .004 & .284 & .980 & .014 & .770 & 1 & .004 & .544 & .996 \\
				250 & .008 & .916 & 1 & .024  & 1 &  1 & .016 & .992 & 1  \\
				400 & .008 & .994 & 1 & .020 & 1 & 1 & .008 & 1 & 1 \\ 
				\hline
				\multicolumn{10}{c}{ Linear-Circular regression. Von Mises errors}\\
				\hline
				&  \multicolumn{3}{c}{$cv/4$}&  \multicolumn{3}{c}{$cv$}&  \multicolumn{3}{c}{$4cv$}\\
				\cmidrule(l){1-1} \cmidrule(l){2-4} \cmidrule(l){5-7} \cmidrule(l){8-10}
				$n$ & $\beta=0$ & $\beta=.5$ & $\beta=1$  & $\beta=0$ & $\beta=.5$ & $\beta=1$ & $\beta=0$ & $\beta=.5$ & $\beta=1$ \\
				50 & .006 & .204 & .934 & .018 & .260 & .512 & .014 & .252 & .504  \\
				100 & .008 & .494 & 1 & .016 & .570 & .906 & .010 & .560 & .900 \\
				250 & .006 & .928 & 1 & .020 & .970 & 1 & .014 & .962 & 1 \\
				400 & .008 & .990 & 1 & .016 & .994 & 1 & .008 & .994 & 1 \\
				\hline
				\multicolumn{10}{c}{ Circular-Circular regression. Von Mises errors}\\
				\hline
				&  \multicolumn{3}{c}{$4cv$}&  \multicolumn{3}{c}{$cv$}&  \multicolumn{3}{c}{$cv/8$}\\
				\cmidrule(l){1-1} \cmidrule(l){2-4} \cmidrule(l){5-7} \cmidrule(l){8-10}
				$n$ & $\beta=0$ & $\beta=.35$ & $\beta=.5$ & $\beta=0$ & $\beta=.35$ & $\beta=.5$ & $\beta=0$ & $\beta=.35$ & $\beta=.5$  \\
				50 &  .008 & .068 & .198 & .026 & .186 & .448 & .010 & .106 & .266  \\ 
				100 & .012 & .268 & .736  & .014 & .484 & .902 & .004 & .242 & .622 \\
				250 &  .008 & .918 & 1 & .008 & .964 & 1 & .008 & .784 & .998  \\ 
				400 & .014 & 1 & 1 & .020 & 1 & 1 & .010 & .990 & 1 \\
				\hline
		\end{tabular}}
		
	\end{center}
\end{table}

\begin{table}[htbp]
	\begin{center}
		\caption{Percentages of rejection (for $\alpha=.10$) obtained with the ANCOVA tests using the smoothing parameters obtained by cross-validation. Results for the first value of $\beta$ show empirical size, whereas results for the other values of $\beta$ show empirical power.}
		\label{table:ANCOVA_010}
		\footnotesize
		\begin{tabular}{cccccccc}
			\hline
			\multicolumn{8}{c}{ Circular-Linear regression. $\chi^2$ calibration. Normal errors}\\
			\hline
			& & \multicolumn{3}{c}{Equality}&  \multicolumn{3}{c}{Parallelism}\\
			\cmidrule(l){1-2} \cmidrule(l){3-5} \cmidrule(l){6-8} 
			$n_1$ & $n_2$  & $\beta=1$ & $\beta=1.5$  & $\beta=1.75$ & $\beta=1$ & $\beta=1.5$  & $\beta=1.75$  \\
			50  & 50  & .090 & .646 & .952 & .094 & .698 & .966  \\
			50  & 100 & .090 & .822 & .992 & .104 & .826 & .992  \\
			100 & 100 & .116 & .968 &    1 & .096 & .972 &    1  \\
			100 & 250 & .102 & .998 & 1 & .104 & .992 & 1 \\
			250 & 250 & .090 & 1 & 1 & .100 & 1 & 1 \\
			\hline
			\multicolumn{8}{c}{ Circular-Linear regression. Bootstrap calibration. Exponential errors}\\
			\hline
			& & \multicolumn{3}{c}{Equality}&  \multicolumn{3}{c}{Parallelism}\\
			\cmidrule(l){1-2} \cmidrule(l){3-5} \cmidrule(l){6-8} 
			$n_1$ & $n_2$  & $\beta=1$ & $\beta=1.5$  & $\beta=1.75$ & $\beta=1$ & $\beta=1.5$  & $\beta=1.75$  \\
			50  & 50  & .102 & .844 &  .994 & .098 & .192 & .326 \\
			50  & 100 & .096 & .922 &     1 & .096 & .202 & .332 \\
			100 & 100 & .106 & .996 &     1 & .084 & .672 & .854  \\
			100 & 250 & .110 & 	  1 &     1 & .106 &    1 &    1  \\
			250 & 250 & .098 &    1 &     1 & .126 &    1 &    1 \\
			\hline
			\multicolumn{8}{c}{ Linear-Circular regression. Von Mises errors}\\
			\hline
			& & \multicolumn{3}{c}{Equality}&  \multicolumn{3}{c}{Parallelism}\\
			\cmidrule(l){1-2} \cmidrule(l){3-5} \cmidrule(l){6-8} 
			$n_1$ & $n_2$ & $\beta=2$ & $\beta=1.75$ & $\beta=1.5$  & $\beta=2$ & $\beta=1.75$ & $\beta=1.5$ \\
			50  & 50  & .104 & .420 & .950 & .114 & .492 & .964 \\
			50  & 100 & .108 & .570 & .988 & .114 & .622 & .996 \\
			100 & 100 & .114 & .760 &    1 & .120 & .764 &    1 \\
			100 & 250 & .110 & .860 &    1 & .092 & .908 &    1 \\
			250 & 250 & .098 & .988 &    1 & .106 & .998 &    1 \\
			\hline
			\multicolumn{8}{c}{ Circular-Circular regression. Von Mises errors}\\
			\hline
			& & \multicolumn{3}{c}{Equality}&  \multicolumn{3}{c}{Parallelism}\\
			\cmidrule(l){1-2} \cmidrule(l){3-5} \cmidrule(l){6-8} 
			$n_1$ & $n_2$ & $\beta=2$ & $\beta=2.5$ & $\beta=3$  & $\beta=2$ & $\beta=2.5$ & $\beta=3$ \\
			50  & 50  & .102 & .570 & .994 & .118 & .518 & .982 \\
			50  & 100 &	.124 & .658 &    1 & .126 & .684 & .996 \\
			100 & 100 & .104 & .902 &    1 & .112 & .898 &    1 \\
			100 & 250 & .128 & .962 &    1 & .126 & .976 &    1 \\
			250 & 250 & .122 &    1 &    1 & .094 & .998 &    1 \\
			\hline
		\end{tabular} 
		
	\end{center}
\end{table}

\begin{table}[htbp]
	\begin{center}
		\caption{Percentages of rejection (for $\alpha=.01$) obtained with the ANCOVA tests using the smoothing parameters obtained by cross-validation. Results for the first value of $\beta$ show empirical size, whereas results for the other values of $\beta$ show empirical power.}
		\label{table:ANCOVA_001}
		\footnotesize
		\begin{tabular}{cccccccc}
			\hline
			\multicolumn{8}{c}{ Circular-Linear regression. $\chi^2$ calibration. Normal errors }\\
			\hline
			& & \multicolumn{3}{c}{Equality}&  \multicolumn{3}{c}{Parallelism}\\
			\cmidrule(l){1-2} \cmidrule(l){3-5} \cmidrule(l){6-8} 
			$n_1$ & $n_2$  & $\beta=1$ & $\beta=1.5$  & $\beta=1.75$ & $\beta=1$ & $\beta=1.5$  & $\beta=1.75$  \\
			50  & 50  & .014 & .310 & .786 & .018 & .350 & .816  \\
			50  & 100 & .006 & .478 & .926 & .008 & .504 & .932  \\
			100 & 100 & .020 & .766 & 1 & .014 & .834 & 1 \\
			100 & 250 & .020 & .944 & 1 & .014 & .960 & 1 \\
			250 & 250 & .012 & 1 & 1 & .012 & 1 & 1 \\
			\hline
			\multicolumn{8}{c}{ Circular-Linear regression. Bootstrap calibration. Exponential errors}\\
			\hline
			& & \multicolumn{3}{c}{Equality}&  \multicolumn{3}{c}{Parallelism}\\
			\cmidrule(l){1-2} \cmidrule(l){3-5} \cmidrule(l){6-8} 
			$n_1$ & $n_2$  & $\beta=1$ & $\beta=1.5$  & $\beta=1.75$ & $\beta=1$ & $\beta=1.5$  & $\beta=1.75$  \\
			50  & 50  & .020 & .482 & .900 & .036 & .040 & .090  \\
			50  & 100 & .018 & .644 & .964 & .008 & .050 & .074 \\
			100 & 100 & .006 & .946 &    1 & .026 & .396 & .618 \\
			100 & 250 & .014 & .996 &    1 & .010 & .992 &    1 \\
			250 & 250 & .004 &    1 &    1 & .006 &    1 &    1 \\
			\hline
			\multicolumn{8}{c}{ Linear-Circular regression. Von Mises errors}\\
			\hline
			& & \multicolumn{3}{c}{Equality}&  \multicolumn{3}{c}{Parallelism}\\
			\cmidrule(l){1-2} \cmidrule(l){3-5} \cmidrule(l){6-8} 
			$n_1$ & $n_2$ & $\beta=2$ & $\beta=1.75$ & $\beta=1.5$  & $\beta=2$ & $\beta=1.75$ & $\beta=1.5$ \\
			50  & 50  & .006 & .106 & .752 & .006 & .190 & .814  \\
			50  & 100 & .012 & .226 & .914 & .016 & .274 & .938 \\
			100 & 100 & .010 & .392 & .994 & .022 & .408 & .994  \\
			100 & 250 & .006 & .622 &    1 & .012 & .644 &    1 \\
			250 & 250 & .010 & .890 &    1 & .014 & .894 &    1 \\
			\hline
			\multicolumn{8}{c}{ Circular-Circular regression. Von Mises errors}\\
			\hline
			& & \multicolumn{3}{c}{Equality}&  \multicolumn{3}{c}{Parallelism}\\
			\cmidrule(l){1-2} \cmidrule(l){3-5} \cmidrule(l){6-8} 
			$n_1$ & $n_2$ & $\beta=2$ & $\beta=2.5$ & $\beta=3$  & $\beta=2$ & $\beta=2.5$ & $\beta=3$ \\
			50  & 50  & .006 & .184 & .866 & .012 & .172 & .852 \\
			50  & 100 &	.012 & .244 & .970 & .012 & .310 & .982  \\
			100 & 100 & .006 & .646 &    1 & .008 & .644 &    1 \\
			100 & 250 & .012 & .834 &    1 & .022 & .870 &    1 \\
			250 & 250 & .014 &    1 &    1 & .006 & .996 &    1 \\
			\hline
		\end{tabular} 
	\end{center}
\end{table}

\begin{table}[htbp]
	\begin{center}
		\caption{Percentages of rejection (for $\alpha=.05$) obtained with the two versions of the no-effect test for circular-linear regression using different smoothing parameters. The shifted and scaled $\chi^2$ calibrated test is applied to exponential errors and the bootstrap version is applied to normal errors. Results for the first value of $\beta$ show empirical size, whereas results for the other values of $\beta$ show empirical power.}
		\label{table:noeffect_switched_errors}
		\resizebox{\textwidth}{!}{
			\footnotesize
			\begin{tabular}{cccccccccc}
				\hline
				\multicolumn{10}{c}{ Circular-Linear regression. $\chi^2$ calibration. Exponential errors}\\
				\hline
				&  \multicolumn{3}{c}{$4cv$}&  \multicolumn{3}{c}{$cv$}&  \multicolumn{3}{c}{$cv/8$}\\
				\cmidrule(l){1-1} \cmidrule(l){2-4} \cmidrule(l){5-7} \cmidrule(l){8-10}
				$n$ & $\beta=0$ & $\beta=.2$ & $\beta=.3$  & $\beta=0$ & $\beta=.2$ & $\beta=.3$ & $\beta=0$ & $\beta=.2$ & $\beta=.3$ \\
				50  & .030 &	.344 &	.920 &	.106 &	.614 &	.992 &	.072 &	.256 &	.828 \\
				100 & .052 &	.858 &	   1 &	.110 &	.920 &	   1 &	.064 &	.598 &	.998 \\ 
				250 & .074 &	   1 &	   1 &	.112 &	   1 &	   1 &	.046 &	.986 &	   1 \\
				400 & .074 &	   1 &	   1 &	.098 &     1 &	   1 &	.046 &	   1 &	   1 \\
				\hline
				\multicolumn{10}{c}{ Circular-Linear regression. Bootstrap calibration. Normal errors}\\
				\hline
				&  \multicolumn{3}{c}{$4cv$}&  \multicolumn{3}{c}{$cv$}&  \multicolumn{3}{c}{$cv/8$}\\
				\cmidrule(l){1-1} \cmidrule(l){2-4} \cmidrule(l){5-7} \cmidrule(l){8-10}
				$n$ & $\beta=0$ & $\beta=.2$ & $\beta=.3$  & $\beta=0$ & $\beta=.2$ & $\beta=.3$ & $\beta=0$ & $\beta=.2$ & $\beta=.3$ \\
				50  & .070 & .236 & .858 &	.104 &	.412 &	.962 &	.060 &	.198 &	.582  \\
				100 & .046 & .578 &	.998 &	.094 &	.782 &	   1 &	.052 &	.292 &	.954 \\
				250 & .060 & .984 &	   1 &	.100 &	.996 &	   1 &	.054 &	.862 &	   1 \\  
				400 & .050 & .998 &	   1 &	.078 &	   1 &	   1 &	.044 &	.988 &	   1 \\
				\hline
			\end{tabular}
		}
	\end{center}
\end{table}

\begin{table}[htbp]
	
	\begin{center}
		\caption{Percentages of rejection (for $\alpha=.05$) obtained with the two versions of the ANCOVA tests for circular-linear regression. The shifted and scaled $\chi^2$ calibrated tests are applied to exponential errors and the bootstrap versions are applied to normal errors. Results for the first value of $\beta$ show empirical size, whereas results for the other values of $\beta$ show empirical power.}
		\label{table:ANCOVA_errors_switched}
		\footnotesize
		\begin{tabular}{cccccccc}
			\hline
			\multicolumn{8}{c}{ Circular-Linear regression. $\chi^2$ calibration. Exponential errors}\\
			\hline
			& & \multicolumn{3}{c}{Equality}&  \multicolumn{3}{c}{Parallelism}\\
			\cmidrule(l){1-2} \cmidrule(l){3-5} \cmidrule(l){6-8} 
			$n_1$ & $n_2$  & $\beta=1$ & $\beta=1.5$  & $\beta=1.75$ & $\beta=1$ & $\beta=1.5$  & $\beta=1.75$  \\
			50  & 50  & .074 &	.700 &	.984 &	.052 &	.768 &	.984 \\
			50  & 100 & .076 &	.880 &	.998 &	.060 &	.894 &	   1 \\
			100 & 100 & .056 &	.986 &	   1 &	.064 &	.996 &	   1 \\
			100 & 250 & .078 &	   1 &	   1 &	.064 &	   1 &	   1 \\
			250 & 250 & .082 &	   1 &	   1 &	.072 &	   1 &	   1 \\
			\hline
			\multicolumn{8}{c}{ Circular-Linear regression. Bootstrap calibration. Normal errors}\\
			\hline
			& & \multicolumn{3}{c}{Equality}&  \multicolumn{3}{c}{Parallelism}\\
			\cmidrule(l){1-2} \cmidrule(l){3-5} \cmidrule(l){6-8} 
			$n_1$ & $n_2$  & $\beta=1$ & $\beta=1.5$  & $\beta=1.75$ & $\beta=1$ & $\beta=1.5$  & $\beta=1.75$  \\
			50  & 50  &	.048 &	.524 &	.898 &	.042 &	.086 &	.136 \\
			50  & 100 & .044 &	.650 &	.974 &	.050 &	.070 &	.142 \\
			100 & 100 & .038 &	.914 &	   1 &	.040 &	.464 &	.694 \\
			100 & 250 & .040 &	.986 &	   1 &	.044 &	.992 &	   1 \\
			250 & 250 & .052 &  	 1 &     1 &	.046 &	   1 &	   1 \\
			\hline
		\end{tabular} 
	\end{center}
\end{table}

\begin{table}[htbp]
	
	\begin{center}
		\caption{Percentages of rejection (for $\alpha=.05$) obtained with the ANCOVA tests using the smoothing parameters which undersmooth the regression curves ($4cv$ for circular predictors, $cv/4$ for real-valued predictors). Results for the first value of $\beta$ show empirical size, whereas results for the other values of $\beta$ show empirical power.}
		\label{table:ANCOVA_undersmoothed}
		\footnotesize
		\begin{tabular}{cccccccc}
			\hline
			\multicolumn{8}{c}{ Circular-Linear regression. $\chi^2$ calibration. Normal errors}\\
			\hline
			& & \multicolumn{3}{c}{Equality}&  \multicolumn{3}{c}{Parallelism}\\
			\cmidrule(l){1-2} \cmidrule(l){3-5} \cmidrule(l){6-8} 
			$n_1$ & $n_2$  & $\beta=1$ & $\beta=1.5$  & $\beta=1.75$ & $\beta=1$ & $\beta=1.5$  & $\beta=1.75$  \\
			50  & 50  & .048 &  .364 &	.766 &	.052 &	.356 &	.782 \\
			50  & 100 & .058 &	.534 &	.914 &	.044 &	.542 &	.930 \\
			100 & 100 & .030 &	.802 &	.990 &	.062 &	.770 &	.993 \\
			100 & 250 & .054 &	.932 &	   1 &	.050 &	.948 &   	 1 \\
			250 & 250 & .042 &    	1 &    1 &	.062 &	   1 &   	 1 \\
			\hline
			\multicolumn{8}{c}{ Circular-Linear regression. Bootstrap calibration. Exponential errors}\\
			\hline
			& & \multicolumn{3}{c}{Equality}&  \multicolumn{3}{c}{Parallelism}\\
			\cmidrule(l){1-2} \cmidrule(l){3-5} \cmidrule(l){6-8} 
			$n_1$ & $n_2$  & $\beta=1$ & $\beta=1.5$  & $\beta=1.75$ & $\beta=1$ & $\beta=1.5$  & $\beta=1.75$  \\
			50  & 50  & .050 &	.574 &	.908 &	.070 &	.120 &	.218  \\
			50  & 100 & .052 &	.710 &	.962 &	.056 &	.138 &	.208  \\
			100 & 100 & .052 &	.926 &	.998 &	.040 &	.530 &	.784  \\
			100 & 250 & .062 &	.982 &	   1 &	.056 &	.992 &	   1  \\
			250 & 250 & .064 &	   1 &	   1 & 	.062 &	   1 &     1  \\
			\hline
			\multicolumn{8}{c}{ Linear-Circular regression. Von Mises errors}\\
			\hline
			& & \multicolumn{3}{c}{Equality}&  \multicolumn{3}{c}{Parallelism}\\
			\cmidrule(l){1-2} \cmidrule(l){3-5} \cmidrule(l){6-8} 
			$n_1$ & $n_2$ & $\beta=2$ & $\beta=1.75$ & $\beta=1.5$  & $\beta=2$ & $\beta=1.75$ & $\beta=1.5$ \\
			
			50  & 50  &  .062 &	.198 &	.686 &	.076 &	.220 &	.712  \\
			50  & 100 &  .048 &	.262 &	.830 &	.044 &	.260 &	.844  \\
			100 & 100 &  .048 &	.358 &	.928 &	.050 &	.382 &	.938  \\
			100 & 250 &  .054 &	.534 &	.998 &	.072 &	.578 &  .996 \\
			250 & 250 &  .042 &	.842 &	   1 &	.052 &	.866 &	   1 \\
			\hline
			\multicolumn{8}{c}{ Circular-Circular regression. Von Mises errors}\\
			\hline
			& & \multicolumn{3}{c}{Equality}&  \multicolumn{3}{c}{Parallelism}\\
			\cmidrule(l){1-2} \cmidrule(l){3-5} \cmidrule(l){6-8} 
			$n_1$ & $n_2$ & $\beta=2$ & $\beta=2.5$ & $\beta=3$  & $\beta=2$ & $\beta=2.5$ & $\beta=3$ \\
			50  & 50  &  .044 &	.244 &	.790 &	.050 &	.242 &	.822 \\
			50  & 100 &	 .054 &	.374 &	.970 &	.066 &	.354 &	.956 \\
			100 & 100 &  .066 &	.596 &	   1 &	.048 &	.646 &	   1 \\
			100 & 250 &  .062 &	.884 &	   1 &	.060 &	.866 &	   1 \\
			250 & 250 &  .056 &	   1 &	   1 &	.048 &	   1 &	   1 \\
			\hline
		\end{tabular} 
	\end{center}
\end{table}

\begin{table}[htbp]
	\begin{center}
		\caption{Percentages of rejection (for $\alpha=.05$) obtained with the ANCOVA tests using the smoothing parameters which oversmooth the regression curves ($cv/8$ for circular predictors, $4cv$ for real-valued predictors). Results for the first value of $\beta$ show empirical size, whereas results for the other values of $\beta$ show empirical power.}
		\label{table:ANCOVA_oversmoothed}
		
		\footnotesize
		\begin{tabular}{cccccccc}
			\hline
			\multicolumn{8}{c}{ Circular-Linear regression. $\chi^2$ calibration. Normal errors}\\
			\hline
			& & \multicolumn{3}{c}{Equality}&  \multicolumn{3}{c}{Parallelism}\\
			\cmidrule(l){1-2} \cmidrule(l){3-5} \cmidrule(l){6-8} 
			$n_1$ & $n_2$  & $\beta=1$ & $\beta=1.5$  & $\beta=1.75$ & $\beta=1$ & $\beta=1.5$  & $\beta=1.75$  \\
			50  & 50  & .160 &	.460 &	.780 &	.290 &	.648 &	.900 \\
			50  & 100 & .152 &	.534 &	.884 &	.220 &	.714 &	.968 \\
			100 & 100 & .154 &	.846 &	.992 &	.202 &	.912 &	.998 \\
			100 & 250 & .104 &	.962 &	   1 &	.142 &	.982 &	   1 \\
			250 & 250 & .114 &	   1 &	   1 &	.152 &	   1 &     1 \\
			\hline
			\multicolumn{8}{c}{ Circular-Linear regression. Bootstrap calibration. Exponential errors}\\
			\hline
			& & \multicolumn{3}{c}{Equality}&  \multicolumn{3}{c}{Parallelism}\\
			\cmidrule(l){1-2} \cmidrule(l){3-5} \cmidrule(l){6-8} 
			$n_1$ & $n_2$  & $\beta=1$ & $\beta=1.5$  & $\beta=1.75$ & $\beta=1$ & $\beta=1.5$  & $\beta=1.75$  \\
			50  & 50  &	.202 &	.690 &	.890 &	.068 &	.102 &	.170  \\
			50  & 100 & .172 &	.740 &	.978 &	.072 &	.116 &	.158  \\
			100 & 100 & .126 &	.964 &	   1 &	.068 &	.442 &	.608  \\
			100 & 250 & .092 &	.998 &	   1 &	.110 &	   1 &	   1  \\
			250 & 250 & .052 &	   1 &	   1 &	.116 &	   1 &	   1 \\
			\hline
			\multicolumn{8}{c}{ Linear-Circular regression. Von Mises errors}\\
			\hline
			& & \multicolumn{3}{c}{Equality}&  \multicolumn{3}{c}{Parallelism}\\
			\cmidrule(l){1-2} \cmidrule(l){3-5} \cmidrule(l){6-8} 
			$n_1$ & $n_2$ & $\beta=2$ & $\beta=1.75$ & $\beta=1.5$  & $\beta=2$ & $\beta=1.75$ & $\beta=1.5$ \\
			50  & 50  &  .122 &	.440 &	.912 &	.132 &	.486 &	.956  \\
			50  & 100 &  .076 &	.614 &	.994 &	.120 &	.642 &	.992  \\
			100 & 100 &  .098 &	.694 &	   1 &	.088 &	.780 &	   1  \\
			100 & 250 &  .064 &	.892 &     1 &	.102 & 	.920 &	   1  \\
			250 & 250 &  .084 &	.984 &	   1 &	.100 &	.992 &	   1 \\
			\hline
			\multicolumn{8}{c}{ Circular-Circular regression. Von Mises errors}\\
			\hline
			& & \multicolumn{3}{c}{Equality}&  \multicolumn{3}{c}{Parallelism}\\
			\cmidrule(l){1-2} \cmidrule(l){3-5} \cmidrule(l){6-8} 
			$n_1$ & $n_2$ & $\beta=2$ & $\beta=2.5$ & $\beta=3$  & $\beta=2$ & $\beta=2.5$ & $\beta=3$ \\
			50  & 50  & .352 &	.688 &	.972 &	.402 &	.706 &	.960  \\
			50  & 100 &	.322 &	.608 &	.980 &	.390 &	.674 &	.984  \\
			100 & 100 & .294 &	.886 &	   1 &	.372 &	.896 &	   1  \\
			100 & 250 & .214 &	.824 &	   1 &	.246 &	.846 &	   1  \\
			250 & 250 & .212 &	.998 &	   1 &	.248 &	.996 &	   1  \\
			\hline
		\end{tabular} 
	\end{center}
\end{table}

\clearpage
\bibliographystyle{apalike}
\bibliography{biblio}

\end{document}